\documentclass{IEEEtran}
\usepackage{cite}
\usepackage{amsmath,amssymb,amsfonts}
\usepackage{algorithmic}
\usepackage{graphicx}
\usepackage{textcomp}
\usepackage{color}
\usepackage{makecell}
\newtheorem{theorem}{\textbf{Theorem}}

\newtheorem{definition}{\textbf{Definition}}
\newtheorem{lemma}{\textbf{Lemma}}
\newtheorem{proposition}{\textbf{Proposition}}

\newtheorem{remark}{\textbf{Remark}}

\usepackage{epstopdf}

\allowdisplaybreaks

\def\BibTeX{{\rm B\kern-.05em{\sc i\kern-.025em b}\kern-.08em
    T\kern-.1667em\lower.7ex\hbox{E}\kern-.125emX}}
\begin{document}
\title{Constructive continuous and sampled-data control of nonstandard singularly perturbed parabolic systems}
\author{Pengfei Wang,  Emilia Fridman, \IEEEmembership{Fellow, IEEE}
\thanks{This work was supported by National Natural Science Foundation of China under Grant (grant no. 62603329), Azrieli International Postdoctoral Fellowship, Israel Science Foundation (grant no. 466/24),  the ISF-NSFC joint research program (grant no. 3054/23),  Chana and Heinrich Manderman Chair on System Control at Tel Aviv University. Pengfei Wang's work was done when he was with Tel Aviv University.}
\thanks{P. Wang is with School of Robotics and Automation, Nanjing University, China, and also with
the School of Electrical Computer \& Engineering, Tel-Aviv University, Tel-Aviv, Israel (e-mail: wangpengfei1156@hotmail.com).}
\thanks{E. Fridman is with
the School of Electrical Computer \& Engineering, Tel-Aviv University, Tel-Aviv, Israel (e-mail: emilia@tauex.tau.ac.il).}
}

\maketitle

\begin{abstract}
The existing results on control of singularly perturbed (SP) PDE systems are all qualitative with stability guaranteed for small values of SP parameter $\varepsilon$, whereas control design is confined to stable fast systems. 
In the present paper we study the finite-dimensional output-feedback boundary stabilization of SP parabolic systems with Neumann actuators and nonlocal sensors acting on both the slow and fast dynamics. 
This configuration allows us to treat not only unstable fast systems, but even a larger class of nonstandard SP systems (where the fast dynamics with $\varepsilon=0$ cannot be solved with respect to the fast state). We design a finite-dimensional controller based on the slow  descriptor PDE system. We consider the continuous-time control and, for the first time, (even for the state-feedback for SP PDEs) its sampled-data implementation. For the latter, we introduce a multi-rate sampled-data framework in which the slow and fast variables can be sampled asynchronously. We consider time-triggered control input sampling and event-triggered output sampling. We prove the avoidance of Zeno behavior and derive LMI conditions for finding upper bounds on the observer dimension, sampling control intervals and $\varepsilon$ that preserve the stability.
Our results include a new finite-dimensional controller design for the parabolic-elliptic equation. Numerical simulations demonstrate the efficiency of the proposed method. 
\end{abstract}

\begin{IEEEkeywords}
Distributed parameter systems, singularly perturbed systems, boundary control, sampled-data control, event-triggered control.
\end{IEEEkeywords}

\section{Introduction}
Singularly perturbed (SP) parabolic systems have attracted significant attention due to their broad applications in engineering and ecology, such as vegetation-water interactions \cite{doelman2022slow}, savanna grass-woodland dynamics \cite{van2003effects}, and thermal fluid convection \cite{vazquez2006explicit}. These systems are characterized by two different time scales arising from a small SP parameter, which significantly influences the performance.

The stability of infinite-dimensional coupled systems with different time scales has been extensively studied in recent years. Examples include coupled ODE with hyperbolic system \cite{tang2017stability,romano2026stability,romano2026passivity}, with wave equation \cite{cerpa2019singular}, and with KdV equation \cite{marx2024singularTAC}, as well as coupled hyperbolic systems \cite{tang2015singular,tang2015tikhonov}, coupled heat-wave systems \cite{arias2025stability}, and coupled infinite-dimensional systems  \cite{marx2024singular}.  
In \cite{romano2026stability,romano2026passivity,cerpa2019singular,marx2024singularTAC,tang2015singular,tang2015tikhonov,arias2025stability,marx2024singular}, stability was established for sufficiently small values of the SP parameter $\varepsilon>0$.
{\color{black}The counterexample in \cite{tang2017stability} shows that the full SP system can be unstable for every $\varepsilon>0$ even when its reduced and boundary-layer systems are exponentially stable, due to a transport-induced delay mechanism in the hyperbolic PDE.  
Such a transport-induced instability does not arise in the class of SP parabolic systems that we study in the present paper (see Remark \ref{remark5}). Controller design for SP parabolic systems was suggested in \cite{vazquez2009boundary,parada2024boundary} using backstepping under the assumption of stable fast systems, where the stability results were qualitative with stability guaranteed for sufficiently small $\varepsilon$. In the recent paper \cite{romano2026stability}, stabilization of a coupled ODE-hyperbolic system was studied under the assumption of a stable fast PDE system, and the resulting stability guarantee was also qualitative.}

For SP ODE systems, Khalil \cite{khalil1989feedback} introduced a design for the nonstandard systems, where differently from the standard systems studied in \cite{kokotovic1986singular} the fast ODE with $\varepsilon=0$ cannot be solved with respect to the fast state. The control was designed on the basis of the slow descriptor system that corresponded to SP system with $\varepsilon=0$ and the fast system. The result was qualitative guaranteeing stability for small enough $\varepsilon>0$. 
Constructive sampled-data state-feedback control of nonstandard SP ODE systems was suggested in \cite{fridman2006robust} via LMI approach to SP systems introduced in \cite{fridman2002effects} with quantitative bounds on $\varepsilon$ and the sampling interval. The objective of our paper is to provide a constructive controller design for SP PDE systems, including nonstandard cases, together with its sampled-data implementation.

When $\varepsilon=0$, SP parabolic systems reduce to slow descriptor systems 
characterized by coupled parabolic-elliptic equations. 
{\color{black}Such systems can also be viewed in the framework of infinite-dimensional differential-algebraic equations (DAEs). The solvability, Weierstra{\ss} forms, and index concepts for infinite-dimensional DAEs have been analyzed in \cite{jacob2022solvability,erbay2024weierstrass,erbay2024index}.}
The boundary stabilization of linear parabolic-elliptic systems has recently been studied in \cite{morris2025boundary} by backstepping and in \cite{parada2020feedback,wang2024finite} by the modal decomposition method, where the system is assumed to be standard (meaning that the state of elliptic equation can be expressed as a Fredholm operator of the parabolic equation state and then substituted into parabolic equation). The latter allows controller design and stability analysis similar to that for heat equations.

Due to the advances of digital implementations, sampled-data control of PDE systems has attracted great attention. Time-triggered sampled-data finite-dimensional controllers for parabolic PDEs were suggested in \cite{fridman2012robust,fridman2013sampled} for distributed static output feedback control, in \cite{karafyllis2018sampled} for boundary state-feedback and in \cite{katz2022delayed,katzasampled,katz2020boundary} for observer-based control. 
Event-triggered control of parabolic PDEs was studied in \cite{espitia2021event,rathnayake2025performance} for boundary state-feedback, in \cite{selivanov2016distributed,kang2021event} for distributed static output control, and in \cite{rathnayake2021observer,yuan2025output,katz2020boundary} for boundary output-feedback. 
In  \cite{rathnayake2026global,somathilake2025output,wang2022event}, the event-triggered control for coupled hyperbolic PDEs was explored. 
{\color{black}For time-triggered sampled-data control of PDEs, the time-delay approach employing combinations of Lyapunov functionals with Halanay's inequality appears to be an efficient tool, as introduced in  \cite{fridman2012robust} for spatial decomposition method and in \cite{katz2022delayed} for modal decomposition. 
However, sampled-data control of SP PDE systems remains an open problem. 
A key challenge lies in handling multi-rate sampled-data design 
(slow sampling for slow variables and fast sampling for fast variables), which is motivated by different time scales of the slow and fast dynamics (see \cite{djemai1999digital,fridman2006robust,chen2020multi} for SP ODEs). The multi-rate samplings of the outputs make the method based on Halanay's inequality, as in \cite{katz2022delayed,katzasampled} for heat equations, inapplicable due to the presence of $\varepsilon$ (see Remark \ref{remark3.1} below). }

In this paper, we suggest a finite-dimensional output-feedback boundary control for SP parabolic systems. 
We consider Neumann actuators and nonlocal sensors in both slow and fast dynamics.  
The controller is designed based on the slow (descriptor) PDE system. 
We consider the continuous-time and, for the first time (even for the state-feedback for PDEs), its sampled-data implementation.
For the latter, we introduce a multi-rate sampled-data framework in which the slow and fast variables can be sampled asynchronously. 
We provide LMI conditions for finding the quantitative upper bound on $\varepsilon$, observer dimension, and the maximum admissible controller sampling intervals that preserve the exponential stability. Numerical examples demonstrate the efficiency of the proposed method. 
The main contribution of this paper is summarized as follows:   
\begin{itemize}
  \item We suggest the first finite-dimensional output-feedback boundary stabilization of SP parabolic systems and provide a constructive method for finding a quantitative upper bound on the SP parameter $\varepsilon$. This is in contrast to the existing stability and stabilization results for SP PDEs that are all qualitative (stability is guaranteed for small enough $\varepsilon$). 
  \item We introduce actuators and nonlocal sensors in both the fast and slow dynamics, enabling the stabilization of the nonstandard SP parabolic systems ($\varepsilon>0$) and slow parabolic-elliptic systems ($\varepsilon=0$). 
  \item We study, for the first time, sampled-data controller implementation for SP PDE systems. We consider time-triggered multi-rate input sampling and control signals implemented via Zero-Order Hold (ZOH) devices. To manage with discrete-time measurements, we introduce a Zeno-free multi-rate event-triggered mechanism. 
\end{itemize}
 
Preliminary results on continuous-time state-feedback boundary stabilization  were reported in CDC 2025 \cite{pengfei2025CDC}.

\textit{Notations:} Denote by $L^{2}(0,1)$ the space of all square integrable functions $f: [0,1]\rightarrow \mathbb{R}$ with inner product $\langle f,g \rangle=\int_{0}^{1}f(x)g(x)\mathrm{d}x$ and induced norm $\|f\|_{L^{2}}^{2}=\langle f,f \rangle$. 
The Euclidean norm is denoted by  $|\cdot|$. For  $P\in \mathbb{R}^{n\times n}$, $P>0$ means that $P$ is symmetric and positive definite. The symmetric elements of a symmetric matrix will be denoted by $*$. For $0<P\in \mathbb{R}^{n\times n}$ and  $x\in \mathbb{R}^{n}$, we write {$|x|^{2}_{P}=x^{\mathrm{T}}Px$}. Denote by {$\mathbb{N}$}  the set of positive integers and {$\mathbb{Z}_+$} the set of nonnegative integers. 
For $m\in \mathbb{N}$ and $\varepsilon \geq 0$, let $E^{\varepsilon}_m=\mathrm{diag}\{I_m, \varepsilon I_m\}$ and $E_m=E^{0}_m$, where $I_m$ means the $m$-dimensional identity matrix. For simplicity, we let $I$ denote an  identity matrix of appropriate dimension and $\mathcal{I}$ be the identity operator on $L^2(0,1)$. {\color{black}For vectors $x_1,\dots,x_m$ of appropriate dimension, we denote $\mathrm{col}\{x_1,\dots,x_m\}=[x_1^{\mathrm{T}},\dots, x_m^{\mathrm{T}}]^{\mathrm{T}}$. Denote by $O(N^k)$, $k\in \mathbb{N}$ the order of $N^k$ time complexity.}

Define the operator:  
\begin{equation}\label{operatorA}
	\begin{split}
		\mathcal{A}: \mathcal{D}(\mathcal{A}) \subset L^{2}(0,1) \to L^{2}(0,1), ~ \mathcal{A}z= -\partial_{xx} z, \\
		\mathcal{D}(\mathcal{A})=\{z\in H^2(0,1), z'(0)=z'(1)=0 \}.  
	\end{split}
\end{equation}
The eigenvalues of $\mathcal{A}$ are given by  
 \begin{equation}\label{lambdan}
 		\lambda_n=(n-1)^2\pi^2, ~ n\in\mathbb{N},  
 \end{equation}
whereas the corresponding eigenfunctions are as follows:  
\begin{equation}\label{SL3}
	\phi_{1}(x)\equiv 1, ~~\phi_n(x) =\sqrt{2}\cos(\sqrt{\lambda_n}x), ~~ n\geq 2. 
\end{equation}
The eigenfunctions \eqref{SL3} form a complete orthonormal system in $L^{2}(0,1)$.

\section{Continuous-time input and output}\label{sec2} 
To facilitate the presentation, we first summarize the main steps of the controller design and stability analysis. First, in Subsection~\ref{sec2A}, modal decomposition is applied to separate the finite-dimensional dominant modes from the infinite-dimensional residual modes. A finite-dimensional observer and controller are then constructed for the dominant modes in Subsection~\ref{sec2B}, followed by the well-posedness analysis in Subsection~\ref{WP}. The stability of the full infinite-dimensional closed-loop system is analyzed in Subsection~\ref{sectionstability}. LMI design of the controller gain from the slow descriptor system is presented in Subsection~\ref{sec2.5}. Based on the resulting controller, The LMI-based stability conditions for the closed-loop SP system are derived in Subsection~\ref{II-F}.

\subsection{Problem formulation and modal decomposition}\label{sec2A} 
We consider the following SP parabolic system: 
\begin{equation}\label{eq1}
	\begin{split}
	&z_t(x,t)=z_{xx}(x,t) - \rho z(x,t)+\alpha v(x,t),\\
	&\varepsilon v_t(x,t) =v_{xx}(x,t) - \gamma v(x,t)+\beta z(x,t),\\
	&z_x(0,t)=0,~~z_x(1,t)=u_1(t),\\
	&v_x(0,t)=0,~~v_x(1,t)=u_2(t), ~t>0,\\
	&z(x,0)=z_0(x),~~v(x,0)=v_0(x),  ~ x\in (0,1),
\end{split}
\end{equation}
where $\varepsilon>0$, $\rho,\gamma,\alpha,\beta\in\mathbb{R}$ with $\alpha\beta\neq 0$, and $u_1(t)$ and $u_2(t)$ are boundary control inputs to be designed.
We consider the following non-local measurement:
\begin{equation}\label{measurement}
		y(t)= \left[ \int_{0}^{1}c_z(x) z(x,t)\mathrm{d}x, ~ \int_{0}^{1}c_v(x) v(x,t)\mathrm{d}x\right]^{\mathrm{T}},
\end{equation}
where $c_z, c_v \in L^2(0,1)$. 

 Setting $\varepsilon=0$, we have the corresponding slow descriptor system (also called the reduced-order system in \cite{parada2024boundary}):
\begin{equation}\label{eq1slow}
{\color{black}	\begin{split}
	&\bar{z}_t(x,t)=\bar{z}_{xx}(x,t) - \rho \bar{z}(x,t)+\alpha \bar{v}(x,t),\\
	&0 =\bar{v}_{xx}(x,t) - \gamma \bar{v}(x,t)+\beta \bar{z}(x,t),\\
	&\bar{z}_x(0,t)=0,~~\bar{z}_x(1,t)=\bar{u}_1(t),\\
	&\bar{v}_x(0,t)=0,~~\bar{v}_x(1,t)=\bar{u}_2(t), ~t>0,\\
	&\bar{z}(x,0)=\bar{z}_0(x),  ~ x\in (0,1). 
\end{split} }
\end{equation}
Introducing the fast time scale $\tau=\frac{t}{\varepsilon}$, 
the corresponding fast system (also called the boundary-layer system in \cite{parada2024boundary}) is given by
\begin{equation}\label{eq1reducedfaster}
{\color{black}	\begin{split}
	&\tilde{v}_\tau(x,\tau) =\tilde{v}_{xx}(x,\tau)-\gamma \tilde{v}(x,\tau),\\
	&\tilde{v}_x(0,\tau)=0,~~\tilde{v}_x(1,\tau)=\tilde{u}_2(\tau), ~\tau> 0,\\
	&\tilde{v}(x,0)=\tilde{v}_0(x), ~ x\in (0,1). 
\end{split} }
\end{equation}
{\color{black}Here, $(\bar z,\bar v)$ and $\tilde v$ are used to distinguish the states of the reduced slow and fast subsystems from the states $(z,v)$ of the original SP system \eqref{eq1}.} 
The slow system \eqref{eq1slow} as well as the SP system \eqref{eq1} is nonstandard if $\gamma=-\lambda_n$ for some $n$, meaning that the elliptic equation in \eqref{eq1slow} cannot be solved with respect to $v$ \cite{morris2025boundary}. This nonstandard SP system extends the nonstandard SP ODE system studied in \cite{khalil1989feedback} to the PDE case.  The fast system \eqref{eq1reducedfaster} is open-loop stable provided  $\gamma >-\lambda_1$,  implying that the SP system is standard. In this paper, we impose no constraints on $\gamma$.

\begin{remark}
{\color{black}Two boundary actuators acting on the slow and fast dynamics were considered in \cite{vazquez2006explicit,vazquez2009boundary} for velocity and temperature control in thermal-fluid convection loops under the constraint $\gamma>-\lambda_1$ (i.e., the fast system is stable). In \cite{parada2024boundary}, state-feedback boundary stabilization of system \eqref{eq1} was studied using the single input $u_1$ under the same constraint $\gamma>-\lambda_1$. In this paper, we remove this constraint by introducing the control input $u_2$. The second input is not introduced merely to simplify the analysis; rather, it enables the proposed approach to address unstable fast dynamics and nonstandard SP systems. In particular, the assumption $\gamma\neq-\lambda_n$ for all $n\in\mathbb{N}$, imposed in \cite{morris2025boundary,parada2020feedback,wang2024finite} for the descriptor system \eqref{eq1slow}, is not required here. In addition, $u_2$ allows us to handle the case
$(\lambda_{n_1}+\gamma)(\lambda_{n_2}+\gamma)=-\alpha\beta$
for some $n_1,n_2\in\mathbb{N}$, which corresponds to multiple eigenvalues of \eqref{eq1slow} and was excluded in \cite{parada2020feedback}. On the other hand, when $\gamma>-\lambda_1=0$ and
$(\lambda_{n_1}+\gamma)(\lambda_{n_2}+\gamma)\neq-\alpha\beta$
for all $n_1,n_2\in\mathbb{N}$, the single input $u_1$ is sufficient. However, the resulting controller may require a high gain, whereas the additional input $u_2$ provides greater design flexibility and can reduce the control gain, as illustrated by the example in Section~\ref{examples}.}
\end{remark}

{\color{black}
For notational simplicity, in the subsequent unified controller design and stability analysis, we return to the notation $(z,v)$ for the descriptor system \eqref{eq1slow} and $v$ for the fast system \eqref{eq1reducedfaster}.}
 We represent the solution to \eqref{eq1} and \eqref{eq1slow} as
\begin{equation*}
 \begin{split}
 	z(\cdot,t)\overset{L^{2}}{=}\sum_{n=1}^{\infty}z_{n}(t)\phi_n,~
	z_{n}(t)=\langle z(\cdot,t), \phi_n\rangle, \\
	v(\cdot,t)\overset{L^{2}}{=}\sum_{n=1}^{\infty}v_{n}(t)\phi_n,~
	v_{n}(t)=\langle v(\cdot,t), \phi_n\rangle,  
 \end{split}
\end{equation*}
with $\{\phi_n\}_{n=1}^{\infty}$ defined in \eqref{SL3}. 
Differentiating under the integral, integrating by parts, using \eqref{operatorA} and \eqref{lambdan}, we obtain
\begin{equation}\label{observereq8aa}
\begin{split}
&	\dot{z}_{n}(t) =(-{\lambda}_n-\rho)  {z}_{n}(t) + \alpha   {v}_n(t) + b_{n}  { u}_1(t),  \\
&\varepsilon \dot{v}_n(t) = (-{\lambda}_n-\gamma) {v}_n(t) +\beta {z}_n(t)+ b_{n} {u}_2(t),\\
&	{z}_{n}(0)=\langle z_0,\phi_{n}\rangle, ~~{v}_{n}(0)=\langle v_0,\phi_{n}\rangle,~~ n\in\mathbb{N},
	\end{split}
\end{equation}
where $b_n=\phi_n(1)$.  Let 
\begin{equation}\label{notations11}
	\begin{split}
	&	A_n= \left[\begin{array}{ccc}
		 - \lambda_n-\rho & \alpha\\
		\beta & -\lambda_n-\gamma
		\end{array}\right], ~   B_n = \left[\begin{array}{cc}
		b_n& 0\\
		0& b_n
		\end{array}\right],\\
	& X_n(t) = [z_n(t), v_n(t) ]^{\mathrm{T}}, ~~ 	{u}(t) =[{u}_1(t) , {u}_2(t)]^{\mathrm{T}}.  
	\end{split}
\end{equation}
We represent \eqref{observereq8aa} as
\begin{equation}\label{studiedsystem2.4}
E_{1}^{\varepsilon}\dot{X}_{n}(t) =A_{n}{X}_{n}(t) + B_{n} {u}(t), ~~ n\in\mathbb{N}.
\end{equation} 
The Fourier coefficients for the fast system \eqref{eq1reducedfaster} satisfy
\begin{equation}\label{fastdecriptor}
\begin{split}
&\dot {{v}}_n(t) = (-{\lambda}_n-\gamma) {v}_n(t) + b_{n} {u}_2(t),\\
&{v}_{n}(0)=\langle v_0,\phi_{n}\rangle, ~~n\in\mathbb{N}.  
	\end{split}
\end{equation}

\subsection{Finite-dimensional controller construction}\label{sec2B}
Let $\delta>0$ be a desired decay rate. 
Since $\lim_{n\to\infty} {\lambda}_n=\infty$, there exists some $N_0\in\mathbb{N}$ such that  
\begin{subequations}\label{NNN}
\begin{align}
&\lambda_{f,n}+\delta <0, ~ \lambda_{f,n}:= -\lambda_n-\gamma, ~n> N_0, \label{fasteigenvalues}\\
&\lambda_{s,n}+\delta<0,  ~ \lambda_{s,n}:= -\lambda_n-\rho + \frac{\alpha\beta}{\lambda_n+\gamma},  ~ n> N_0,    \label{sloweigenvalues}
\end{align} 
\end{subequations}
where ${\lambda}_n$ are defined in \eqref{lambdan}. 
Here, $\lambda_{f,n}$ and $\lambda_{s,n}$ ($n > N_0$) denote the characteristic roots of the fast subsystem \eqref{fastdecriptor} and the slow subsystem \eqref{observereq8aa} with $\varepsilon = 0$, respectively. 
Condition \eqref{fasteigenvalues} (resp. \eqref{sloweigenvalues}) ensures that the modes with $n>N_0$ of the uncontrolled fast subsystem \eqref{fastdecriptor} (resp. the uncontrolled slow subsystem \eqref{observereq8aa} with $\varepsilon=0$) are exponentially stable with a decay rate larger than $\delta$.
Therefore, $N_0$ satisfying \eqref{NNN} characterizes the number of potentially unstable modes, and we design the controller gain based on the first $N_0$ modes. Let   
\begin{align}\label{notationsN0}
&	c_{z,n}=\langle c_z, \phi_n \rangle, ~~~~~~ c_{v,n}=\langle c_v, \phi_n \rangle, ~~n\in\mathbb{N},  \nonumber \\
&{\bf C}_0=\left[\begin{array}{cc}
		C_{z,0} & 0 \\ 
		0  &  C_{v,0}
		\end{array}\right], ~ \begin{array}{cc}
		C_{z,0}=[c_{z,1}, \dots, c_{z,N_0}], \\
		C_{v,0}=[c_{v,1}, \dots, c_{v,N_0}], \\
		\end{array}  \nonumber\\ 
& {\bf A}_0=   \left[\begin{array}{cc}
		 {A}_{\rho,0}  & \alpha I\\ 
		\beta I &  {A}_{\gamma,0}
		\end{array}\right], ~	
		\begin{array}{cc}
		 {A}_{\rho,0} = -\mathrm{diag}\{ \lambda_n \}_{n=1}^{N_0}-\rho I, \\ 
		{A}_{\gamma,0} = -\mathrm{diag}\{ \lambda_n \}_{n=1}^{N_0}-\gamma I,
		\end{array}  \nonumber\\
& {\bf B}_0=\left[\begin{array}{cc}
		B_0  & 0 \\ 
		0  &  B_0
		\end{array}\right],~~~~~~B_0=\mathrm{col}\{b_n\}_{n=1}^{N_0}.       
\end{align}
We will design the controller and observer gains based on the descriptor ODE system $(E_{N_0}, {\bf A}_0, {\bf B}_0)$ governed by 
\begin{equation*}
	E_{N_0}\dot{X}_0 ={\bf A}_0X_0+{\bf B}_0u.
\end{equation*} 
We recall the following definition.
\begin{definition}[\cite{dai1989singular}]
	The descriptor system $(E_{N_0}, {\bf A}_0, {\bf B}_0)$ is said to be
	 stabilizable, if there exists a state feedback $u=-K_0X_0$ such that the closed-loop system $(E_{N_0}, {\bf A}_0-{\bf B}_0K_0)$ governed by 
 \begin{equation}\label{eqclosedstable}
 	E_{N_0}\dot{X}_0 =({\bf A}_0-{\bf B}_0K_0)X_0
 \end{equation}
 is stable, meaning that all roots of $\mathrm{det}(sE_{N_0}-({\bf A}_0-{\bf B}_0K_0))=0$ have negative real parts. 
\end{definition}

System \eqref{eqclosedstable} with $E_{N_0}=\mathrm{diag}\{I_{N_0}, 0_{N_0\times N_0}\}$ is regular (i.e., $\mathrm{det}(sE_{N_0}-({\bf A}_0-{\bf B}_0K_0))$ is not identically zero) and impulse-free (i.e., $\mathrm{deg}(\mathrm{det}(sE_{N_0}-({\bf A}_0-{\bf B}_0K_0)))=N_0$). 
The following lemma was established in the conference version \cite{pengfei2025CDC}.
\begin{lemma}\label{proposition2.1} 
The slow descriptor system	$(E_{N_0}, {\bf A}_0, {\bf B}_0)$ is stabilizable.  
\end{lemma}

Let ${\bf A}^{\delta}_0={\bf A}_0+\delta E_{N_0}$, where $\delta>0$ is the desired decay rate. It is clear that $(E_{N_0}, {\bf A}_0^{\delta}, {\bf B}_0)$ is stabilizable. Since the pair $(A_{\gamma,0}+\delta I, [B_0, 0_{1\times N_0}])$ is controllable, it follows from the proof of Lemma 3.2 in \cite{khalil1989feedback} that there always exists a gain matrix $K_0= (K_{ij})_{2\times 2} \in\mathbb{R}^{2\times 2N_0}$ with  
\begin{equation}\label{A1234}
	\begin{array}{ll}
	{\bf A}_{K_0}=	{ \left[\begin{array}{cc}
		{\bf A}_{11} & {\bf A}_{12}\\
		{\bf A}_{21} & {\bf A}_{22}
		\end{array}\right]}:= {\bf A}_0^{\delta}-{\bf B}_0K_0,  
	\end{array}
\end{equation} 
such that ${\bf A}_{22}$ is Hurwitz and $(E_{N_0}, {\bf A}_{K_0})$ is stable. 
Therefore, by \cite[Theorem 2.1]{xu2006robust}, there exists a matrix 
\begin{equation}\label{eqP_c}
P_c={ \left[ \begin{array}{ccc}
		P_{c1}  & 0 \\
		P_{c2}  & P_{c3}
		\end{array} \right]}	,~~ 0<P_{c1},P_{c3}\in\mathbb{R}^{N_0\times N_0}	,
\end{equation}
such that
	\begin{equation}\label{Lyapunoveq1}
P_c^{\mathrm{T}}({\bf A}_0^{\delta} - {\bf B}_0 K_0) + ({\bf A}_0^{\delta} - {\bf B}_0 K_0)^{\mathrm{T}} P_c  < -\chi_1 I,
\end{equation}	
for small enough $\chi_1>0$. 
\begin{remark}\label{rmk2.2}
 The gain $K_0$ satisfying \eqref{Lyapunoveq1} corresponds to the descriptor ODE system. In Section \ref{sec2.5}, we will give a precise design for $K_0$ (see LMI \eqref{dyLMIa100ss} and expression \eqref{KKss}) that corresponds to the descriptor PDE system (i.e., \eqref{eq225}, \eqref{SPsec5eq1ss}, \eqref{eq224}  with $\varepsilon=0$). 
From \eqref{dyLMIa100ss}, it follows that \eqref{Lyapunoveq1} holds with small enough $\chi_1$. 
	\end{remark}

Let $N\geq N_0$ be the observer dimension. Following \cite{katz2020constructive} for heat equations, we construct the finite-dimensional observer  
\begin{equation}\label{finobserver}
		\hat{z}(x,t)=\sum_{n=1}^{N}\hat{z}_n(t)\phi_{n}(x), ~
		\hat{v}(x,t)=\sum_{n=1}^{N}\hat{v}_n(t)\phi_{n}(x),
\end{equation}
with  $\hat{X}_n(t)=[\hat{z}_n(t), \hat{v}_n(t)]^\mathrm{T}$, $1\leq n\leq N$,  satisfying  
\begin{equation}\label{observerdyeq44}
 \begin{split}
&	E^\varepsilon_1 \dot{\hat{X}}_{n}(t) =A_n \hat{X}_{n}(t) +  B_{n}{u}(t)\\
&	~~~~~ - L_n \left(  \left[ \begin{array}{cc}
		\int_{0}^{1}c_z(x) \hat{z}(x,t)\mathrm{d}x\\
		\int_{0}^{1}c_v(x) \hat{v}(x,t)\mathrm{d}x
		\end{array} \right] -y(t) \right), ~t\geq 0,  \\
&		\hat{X}_{n}(0)=0,  
\end{split}
\end{equation}
where $A_n$, $B_n$ are defined in \eqref{notations11} and $y(t)$ is defined in \eqref{measurement}, $L_{n}= (L_{n}^{(ij)})_{2\times 2} \in\mathbb{R}^{2\times 2}$, $n=1,\dots, N$, are the observer gains to be designed. Following \cite{katz2020constructive,sakawa1983feedback}, we set $L_n=0_{2\times 2}$ for $n=N_0+1,\dots, N$. 
 Assume  
\begin{equation}\label{cneq0}
	c_{z,n}\neq 0, ~~c_{v,n}\neq 0, 	~~n=1,\dots,N_0.  
\end{equation}
\begin{remark}
{\color{black}	Assumption \eqref{cneq0} is a sufficient condition required by the proposed observer design 
when the fast subsystem is unstable ($\gamma<-\lambda_1$) or the system is nonstandard ($\gamma=-\lambda_n$ for some $n\in\mathbb{N}$). 
A similar assumption is used for parabolic PDEs (see, e.g., \cite{katz2022delayed}). When $\gamma>-\lambda_1$, the fast subsystem is exponentially stable and $A_{\gamma,0}$ is invertible. In this case, the fast variable can be eliminated from the descriptor system, and direct measurement of $v$ is not necessary (i.e., one may take $c_v\equiv0$); it suffices that the potentially unstable modes of the resulting reduced slow system are detectable from the measurement of $z$.}
\end{remark}

By Assumption \eqref{cneq0} and arguments similar to the proof of Lemma \ref{proposition2.1}, we can obtain that $(E_{N_0}, ({\bf A}_0^{\delta})^{\mathrm{T}},{\bf C}^{\mathrm{T}}_0)$ is stabilizable. 
We choose $L_{n}= (L_{n}^{(ij)})_{2\times 2}$, $n=1,\dots, L_{N_0}$, such that $(E_{N_0}, ({\bf A}_0^{\delta})^{\mathrm{T}}-{\bf C}_0^{\mathrm{T}}{\bf L}_0^{\mathrm{T}})$ is stable, 
where 
\begin{equation*}
{\bf L}_0=({\bf L}_0^{(ij)})_{2\times 2},~ {\bf L}_0^{(ij)}=	[L_{1}^{(ij)} , \dots , L_{N_0}^{(ij)}], i,j=1,2.
\end{equation*}
From \cite[Corollary 2.2]{xu2006robust} and arguments similar to \cite[Theorem 3.1]{xu2006robust}, we deduce that there exist matrices $0<{\bf P}_o\in\mathbb{R}^{2N_0\times 2N_0}$, ${\bf Q}_o=[{\bf Q}_{o1}, {\bf Q}_{o2} ]$, ${\bf Q}_{o1} \in \mathbb{R}^{N_0\times N_0}$, $0<{\bf Q}_{o2} \in \mathbb{R}^{N_0\times N_0}$, and $Y_o \in  \mathbb{R}^{2N_0\times 2}$ such that   
\begin{equation}\label{observergain11}
 \begin{split}
 &({\bf P}_oE_{N_0}+S_o{\bf Q}_o)^{\mathrm{T}}{\bf A}_0^{\delta} + ({\bf A}_0^{\delta})^{\mathrm{T}} ({\bf P}_oE_{N_0}+S_o{\bf Q}_0) \\
 & - Y_o{\bf C}_0 -{\bf C}_0^{\mathrm{T}}Y_o <0,  
\end{split}
\end{equation}
is feasible, where $S_o=[ 0_{N_0 \times N_0} ,  I_{N_0} ]^{\mathrm{T}}$.  Then the $\varepsilon$-independent gain ${\bf L}_0$ is given by  
\begin{equation}\label{L0L0}
		{\bf L}_0=({\bf P}_oE_{N_0}+S_o{\bf Q}_o)^{-\mathrm{T}}Y_o. 
\end{equation}
From \eqref{observergain11}, we see that the obtained ${\bf L}_0$ in \eqref{L0L0} always guarantees that $A_{\gamma,0}-C_{v,0}{\bf L}_0^{(22)}$ is Hurwitz.

We propose the following finite-dimensional controller:  
\begin{equation}\label{observercontroller}
	\begin{split}
		& u(t)=[u_1(t),u_2(t)]^{\mathrm{T}}=-K_0 \hat{X}_0(t), \\
		& \hat{X}_0=\mathrm{col}\{\hat{z}_1,\dots, \hat{z}_{N_0}, \hat{v}_1,\dots,\hat{v}_{N_0}\},  
	\end{split}
\end{equation}
where $K_0=(K_{ij})_{2\times 2}  \in\mathbb{R}^{2\times 2N_0}$ is subject to \eqref{Lyapunoveq1}.

\subsection{Well-posedness of the closed-loop system}\label{WP}
For the well-posedness of system \eqref{eq1} and \eqref{observerdyeq44} subject to the control law \eqref{observercontroller} with controller gain $K_0$ satisfying \eqref{Lyapunoveq1} and observer gain ${\bf L}_0$ obtained from \eqref{observergain11} and given by \eqref{L0L0}, we introduce the change of variables (see, e.g., \cite{karafyllis2021lyapunov}):
\begin{equation}\label{changeofvariablesdescriptor}
	\begin{split}
	&	{\mathcal{Z}}(x,t)=\bar{z}(x,t)-\psi(x)u_1(t), \\
	&	{\mathcal{V}}(x,t)=\bar{v}(x,t)-\psi(x)u_2(t), \\
	&	\psi(x)=(-1)^k  \frac{1}{\mu}\cos(\mu x), ~\mu=\frac{(2k-1)\pi}{2}, 
	\end{split}
\end{equation}
where $k\in\mathbb{N}$ in $\psi(x)$ is chosen freely.  
Note that the gains $K_0$ and ${\bf L}_0$ are selected independently of $\psi$.  We can choose $k \in \mathbb{N}$ sufficiently large (so that $|\psi|^2_{L^2}$ is small enough) in the well-posedness analysis.

From \eqref{eq1} and \eqref{changeofvariablesdescriptor}, it follows that
\begin{subequations}\label{eq:transformed}
\begin{align}
&{\mathcal{Z}}_t(t) = -\mathcal{A}{\mathcal{Z}}(t) - \rho {\mathcal{Z}}(t) + \alpha {\mathcal{V}}(t)  \notag \\
& \quad\quad~ - \psi(\cdot)[(\mu+\rho)u_1(t) - \alpha u_2(t) + \dot{u}_1(t)], \label{eq:trans1}\\
&\varepsilon {\mathcal{V}}_t(t) = -\mathcal{A}{\mathcal{V}}(t) - \gamma{\mathcal{V}}(t) + \beta{\mathcal{Z}}(t) \notag  \\ 
&\quad\quad ~ - \psi(\cdot)[(\mu+\gamma)u_2(t) - \beta u_1(t)+ \varepsilon \dot{u}_2(t) ], \label{eq:trans2}
\end{align}
\end{subequations}
where $\mathcal{A}$ is defined in \eqref{operatorA}. 
Recall the observer states $\hat{z}, \hat{v}$ given by \eqref{finobserver} with $\hat{z}_n(t), \hat{v}_n(t)$, $n=1,\dots,N$ satisfying \eqref{observerdyeq44}. 
Denote the projection operator $P_N:L^2(0,1)\to\mathbb{R}^N$ by $P_N f = \mathrm{col}\{\langle f,\phi_n\rangle\}_{n=1}^N$ and define
\begin{equation*}
	\begin{split}
&\hat{\mathcal{Z}}_n(t) = \hat{z}_n(t) - \psi_n u_1(t),~
	\hat{\mathcal{V}}_n(t) = \hat{v}_n(t) - \psi_n u_2(t), \\
&\psi_n = \langle \psi, \phi_n \rangle=\frac{b_n}{\lambda_n-\mu^2}\neq 0, \\
&\mathcal{Z}^N(t) = P_N\mathcal{Z}(\cdot,t) ,~ \mathcal{V}^N(t) = P_N\mathcal{V}(\cdot,t),\\
&\hat{\mathcal{Z}}^N = [\hat{\mathcal{Z}}_1,\dots,\hat{\mathcal{Z}}_N]^{\mathrm{T}},~\hat{\mathcal{V}}^N = [\hat{\mathcal{V}}_1,\dots,\hat{\mathcal{V}}_N(t)]^{\mathrm{T}},\\
&\mathcal{C}_z^N\mathcal{Z}=\sum_{n=N+1}^\infty c_{z,n}\mathcal{Z}_n,~ \mathcal{C}_v^N\mathcal{V}=\sum_{n=N+1}^\infty c_{v,n}\mathcal{V}_n, \\
& C_z = \mathrm{row}\{ \langle c_z, \phi_n\rangle \}_{n=1}^{N}, ~C_v = \mathrm{row}\{ \langle c_v, \phi_n\rangle \}_{n=1}^{N}, \\
&\mathcal{E}_z(t) = \mathcal{Z}^N(t) - \hat{\mathcal{Z}}^N(t), ~
\mathcal{E}_v(t) = \mathcal{V}^N(t) - \hat{\mathcal{V}}^N(t), \\
&\mathcal{L}=(\mathcal{L}_{ij})_{2\times 2}, ~\mathcal{L}_{ij}=[ 	({\bf L}_0^{(ij)})^{\mathrm{T}} ,~ 0_{1\times (N-N_0)} ]^{\mathrm{T}}\in \mathbb{R}^{N}. 
	\end{split}
\end{equation*} 
From \eqref{observerdyeq44}, we obtain
\begin{equation}\label{observerdyeq44ob}
 \begin{split}
&	\left[\begin{matrix}
		\dot{\hat{\mathcal{Z}}}^N(t)\\
		\varepsilon \dot{\hat{\mathcal{V}}}^N(t)
	\end{matrix}\right]=\begin{bmatrix} A_\rho & \alpha I \\ \beta I & A_\gamma \end{bmatrix} \left[\begin{matrix}
		{\hat{\mathcal{Z}}}^N(t)\\
		{\hat{\mathcal{V}}}^N(t)
	\end{matrix}\right]+\left[\begin{matrix}
		B & 0\\
		0 & B
	\end{matrix}\right]{u}(t)\\
&\quad\quad\quad\quad 	- \mathcal{L}  \left[ \begin{array}{cc}
		C_z\hat{\mathcal{Z}}^N  - \langle c_z, \mathcal{Z}(t)\rangle - \psi_{c_z,N}^{\infty} u_1(t) \\
		C_v\hat{\mathcal{V}}^N  - \langle c_v, \mathcal{V}(t)\rangle - \psi_{c_c,N}^{\infty} u_2(t)
		\end{array} \right],
\end{split}
\end{equation}
where $\psi_{c_z,N}^{\infty}=\sum_{n=N+1}^{\infty} c_{z,n}\psi_n$ and $\psi_{c_v,N}^{\infty}=\sum_{n=N+1}^{\infty} c_{v,n}\psi_n$. 
By \eqref{eq:transformed} and  \eqref{observerdyeq44ob}, we get the error dynamics
\begin{equation}\label{observer_error_dyn}
\begin{bmatrix}
\dot{\mathcal{E}}_z \\ \varepsilon \dot{\mathcal{E}}_v
\end{bmatrix}
= \begin{bmatrix} A_\rho & \alpha I \\ \beta I & A_\gamma \end{bmatrix}\begin{bmatrix}
\mathcal{E}_z \\ \mathcal{E}_v
\end{bmatrix}
- \mathcal{L}\begin{bmatrix}
C_z\mathcal{E}_z - \mathcal{C}_z^N\mathcal{Z} - \psi_{c_z,N}^{\infty} u_1 \\
C_v\mathcal{E}_v - \mathcal{C}_v^N\mathcal{V} - \psi_{c_v,N}^{\infty} u_2
\end{bmatrix}.
\end{equation}
From \eqref{observercontroller}, it follows
\begin{equation*}
 \begin{split}
&		 \left[\begin{matrix}
		 	u_1(t)\\
		 	u_2(t)
		 \end{matrix}\right]	 = -\mathcal{K} \left[\begin{matrix}
		 	\hat{\mathcal{Z}}^{N}(t)+\psi^{N} u_1(t)\\
		 	\hat{\mathcal{V}}^{N}(t)+\psi^{N} u_2(t)
		 \end{matrix}\right],\\
&	{\mathcal{K}}=({\mathcal{K}}_{ij})_{2\times 2},	~{\mathcal{K}}_{ij}=[{K}_{ij}, 0_{1\times (N-N_0)}], i,j=1,2. 
	\end{split}
\end{equation*}
Solving for $u(t)$ yields
\begin{equation}\label{observercontrollerZV22}
	\begin{split}
&		 \left[\begin{matrix}
		 	u_1(t)\\
		 	u_2(t)
		 \end{matrix}\right] =-\tilde{\mathcal{K}} \left[\begin{matrix}
		 	\hat{\mathcal{Z}}^{N}(t)\\
		 	\hat{\mathcal{V}}^{N}(t)
		 \end{matrix}\right] =-\tilde{\mathcal{K}} \left[\begin{matrix}
		 	P_N\mathcal{Z}(t)-\mathcal{E}_z(t)\\
		 	P_N\mathcal{V}(t)-\mathcal{E}_v(t)
		 \end{matrix}\right],\\
&	\tilde{\mathcal{K}} = (\tilde{\mathcal{K}}_{ij})_{2\times 2} = (I_2 + {\mathcal{K}} \mathrm{diag}\{\psi^{N},\psi^{N}\})^{-1} {\mathcal{K}}.	 
	\end{split}
\end{equation} 
Here, the invertibility of $I_2+{\mathcal{K}}\mathrm{diag}\{\psi^{N},\psi^{N}\}$ can always be guaranteed by a suitable choice of $k$ in $\psi$.

Define $\mathcal{X}_z=(\mathcal{Z},\mathcal{E}_z)\in L^2(0,1)\times\mathbb{R}^N$ and $\mathcal{X}_v=(\mathcal{V},\mathcal{E}_v)\in L^2(0,1)\times\mathbb{R}^N$. Substituting \eqref{observercontrollerZV22} into \eqref{eq:transformed} and \eqref{observer_error_dyn}, we obtain the closed-loop system in the form
\begin{subequations}\label{closedwp2}
\begin{align}
&\dot{\mathcal{X}}_z(t)=\mathcal{A}_{\varepsilon}\mathcal{X}_z(t)+\mathcal{B}_{\varepsilon} \mathcal{X}_v(t),  \label{closedwpz2}\\
&\varepsilon\dot{\mathcal{X}}_v(t)  =\mathcal{M}_{\varepsilon} \mathcal{X}_v(t)+\mathcal{N}_{\varepsilon}\mathcal{X}_z(t), \label{closedwpv2}
\end{align}
\end{subequations}
where 
\begin{equation*}
	\begin{split}
&		\mathcal{A}_{\varepsilon} = \bar{\mathcal{A}}_{\varepsilon}+ ({\mathcal{A}}_{ij}^{\varepsilon})_{2\times 2},~~\mathcal{M}_{\varepsilon} = \bar{\mathcal{M}}_{\varepsilon} + ({\mathcal{M}}_{ij}^{\varepsilon})_{2\times 2},
		\\		 
&	\bar{\mathcal{A}}_{\varepsilon}=\begin{bmatrix} -\mathcal{A} -\rho \mathcal{I}  & 0 \\ 
		0 & 0_{N\times N} 
		\end{bmatrix},~	 
		\bar{\mathcal{M}}_{\varepsilon} = \begin{bmatrix} -\mathcal{A} -\gamma \mathcal{I}  & 0 \\ 
		0 & 0_{N\times N}  
		\end{bmatrix},
	\end{split}
\end{equation*}
with ${\mathcal{A}}_{ij}^{\varepsilon}$, ${\mathcal{M}}_{ij}^{\varepsilon}$, $i,j=1,2$, $\mathcal{B}_{\varepsilon}$, and $\mathcal{N}_{\varepsilon}$ being bounded operators for each $\varepsilon\geq 0$, 
since they either act on the finite-dimensional space $\mathbb{R}^N$ or are bounded operators between $L^2(0,1)$ and $\mathbb{R}^N$ (they involve only multiplication by $\psi$ and bounded linear functionals).
Their explicit forms are omitted due to space constraints.
 
First, for any fixed $\varepsilon>0$, following arguments similar to \cite[Eqs.~(24)-(26)]{katz2021finite}, we have that 
for any initial condition $(\mathcal{X}_{z,0},\mathcal{X}_{v,0})\in \mathcal{D}((\mathcal{A}^{\rm cl}_{\varepsilon})^{\frac{1}{2}})$, where $\mathcal{A}^{\rm cl}_{\varepsilon}=\mathrm{diag}\{\bar{\mathcal{A}}_{\varepsilon}, \frac{1}{\varepsilon}\bar{\mathcal{M}}_{\varepsilon}\}$, there exists a unique classical solution: $(\mathcal{X}_z, \mathcal{X}_v)\in C^1((0,\infty), L^2(0,1)\times \mathbb{R}^{N}\times L^2(0,1)\times \mathbb{R}^{N})$ and $(\mathcal{X}_z(t), \mathcal{X}_v(t))\in \mathcal{D}(\mathcal{A}^{\rm cl}_{\varepsilon})$ for all $t\geq 0$. Applying the transformation \eqref{changeofvariablesdescriptor}, we obtain that for initial data $z_0, v_0 \in H^1(0,1)$, the original system \eqref{eq1} with any fixed $\varepsilon>0$ and the control law \eqref{observercontroller} admits a unique classical solution $z,v \in C([0,\infty), L^2(0,1)) \cap C^1(0,\infty), L^2(0,1)$ and $z(\cdot,t), v(\cdot,t) \in H^2(0,1)$ with $z_x(0,t)=v_x(0,t)=0$, $z_x(1,t)=u_1(t)$, $v_x(1,t)=u_2(t)$ for all $t \ge 0$.

For $\varepsilon=0$, we have $\mathcal{M}_0$ in \eqref{closedwpv2} given by  
\begin{equation}\label{MM}
\begin{split}
& \mathcal{M}_0=(\mathcal{M}_{ij})_{2\times 2},\\
&\mathcal{M}_{11}=-\mathcal{A} -\gamma \mathcal{I}+\tilde{\psi}(\cdot)\tilde{\mathcal{K}}_{22}P_{N},\\
&\mathcal{M}_{12}=	-\tilde{\psi}(\cdot)\tilde{\mathcal{K}}_{22}, ~~\tilde{\psi}=(\mu+\gamma)\psi,\\
&\mathcal{M}_{21}=\mathcal{L}_{22}(\mathcal{C}_{v}^{N}-\psi_{c_v}^N\tilde{\mathcal{K}}_{22}P_N),\\
&\mathcal{M}_{22}=	A_{\gamma}-\mathcal{L}_{22}C_v-\mathcal{L}_{22}\psi_{c_v}^{N}\tilde{\mathcal{K}}_{22}.   
\end{split}
\end{equation}
From \eqref{fasteigenvalues}, we have $\gamma>-\lambda_{N_0+1}+\delta$. 
In the following, we only consider the non-standard case $\gamma=-\lambda_{n_0}$ for some $n_0\in\{1,\dots,N_0\}$. The case $\gamma\neq -\lambda_n$, $n=1,\dots, N_0$, is standard and follows similarly. 
To establish the well-posedness of \eqref{closedwp2} with $\varepsilon=0$, we first prove that $\mathcal{M}_0$ is invertible with a bounded inverse in the following proposition, with the proof given in Appendix \ref{proofpropoWP}. 
\begin{proposition}\label{propoWP}
The operator $\mathcal{M}_0:L^2(0,1)\times\mathbb{R}^N\to L^2(0,1)\times\mathbb{R}^N$ is boundedly invertible.
\end{proposition}

Since $\mathcal{M}_0$ is boundedly invertible, following the well-posedness analysis in \cite{wang2024finite}, we have that for any initial condition $\mathcal{X}_{z,0}\in \mathcal{D}(\mathcal{A}_{0}^{\frac{1}{2}})$, there exists a classical solution $\mathcal{X}_z\in C^1((0,\infty),\mathcal{H}_z)$ and $\mathcal{X}_z(t)\in \mathcal{D}(\mathcal{A}_{0})$ for all $t\geq 0$. Applying the transformation \eqref{changeofvariablesdescriptor}, we obtain that for ${z}_0\in H^1(0,1)$, the original system \eqref{eq1slow} with control law \eqref{observercontroller} admits a unique solution  ${z},{v}\in C([0,\infty),L^2(0,1))\cap C^1((0,\infty),L^{2}(0,1))$ and ${z}(\cdot,t),{v}(\cdot,t)\in H^2(0,1)$ with $z_x(0,t)=v_{x}(0,t)=0$, $z_x(1,t)=u_{1}(t)$, $v_x(1,t)=u_{2}(t)$ for all $t\geq 0$.

\subsection{Stability analysis of the closed-loop system}\label{sectionstability}
In this section, given the $\varepsilon$-independent gains $K_0$ and ${\bf L}_0$, we derive stability conditions for the closed-loop system \eqref{eq1} under the  observer \eqref{finobserver}, \eqref{observerdyeq44} and the controller \eqref{observercontroller} for $\varepsilon\ge 0$.  
Let $e_{n}(t)= [e_{z,n}(t),e_{v,n}(t)]^{\mathrm{T}} =X_{n}(t)-\hat{X}_{n}(t)$. The last term on the right-hand side of \eqref{observerdyeq44} can be rewritten as  
\begin{align}\label{observereq8}
	&	\left[ \begin{array}{cc}
		\int_{0}^{1}c_z(x) \hat{z}(x,t)\mathrm{d}x\\
		\int_{0}^{1}c_v(x) \hat{v}(x,t)\mathrm{d}x
		\end{array} \right] -y(t) =-\sum_{n=1}^{N}C_ne_n(t)-\zeta(t),  \nonumber\\
	&	C_n=\mathrm{diag}\{ c_{z,n}, c_{v,n} \}, ~~ \zeta(t)=[\zeta_{z}(t), \zeta_{v}(t)  ]^{\mathrm{T}},  \nonumber \\
	&	\zeta_{z}=\sum_{n=N+1}^{\infty}c_{z,n} z_n,~~ 
		\zeta_{v}=\sum_{n=N+1}^{\infty}c_{v,n} v_n.  
\end{align}
From the Cauchy-Schwarz inequality, we obtain  
\begin{equation}\label{zetazeta}
	\begin{split}
	&	|\zeta_z(t)|^{2}\leq \|c_z\|^{2}_{N} \sum_{n=N+1}^{\infty} |z_n(t)|^2,\\
&|\zeta_v(t)|^{2}\leq \|c_v\|^{2}_{N} \sum_{n=N+1}^{\infty} |v_n(t)|^2,  
	\end{split}
\end{equation}
where $\|c_z\|^{2}_{N}=\sum_{n=N+1}^{\infty} c_{z,n}^2$ and $\|c_v\|^{2}_{N}=\sum_{n=N+1}^{\infty} c_{v,n}^2$. 
By \eqref{studiedsystem2.4} and \eqref{observerdyeq44}, we have the error equations:  
\begin{subequations}\label{error}
\begin{align}
&  E_1^{\varepsilon} \dot{e}_n(t)=A_n e_n(t) -L_n\left[\sum_{n=1}^{N}C_ne_n(t)+\zeta(t) \right] ,    \label{error1}\\
&  ~~~~~~~~~~~~~~~~~~~~~~~~~~~~~~~~~~~~~~~~~~~ 1\le  n\leq N_0,  \nonumber \\
&  E_1^{\varepsilon} \dot{e}_n=A_n e_n , ~~N_0<n\leq N.  \label{error2}
\end{align}
\end{subequations} 

Recall the notations in \eqref{notations11}, \eqref{notationsN0}, \eqref{observercontroller}, and  introduce  
\begin{equation}\label{notation222}
	\begin{split}
&		 {\bf A}_1=   \left[\begin{array}{cc}
		{A}_{\rho,1}   & \alpha I\\ 
		\beta I &  {A}_{\gamma,1}
		\end{array}\right], 
		\begin{array}{cc}
		{A}_{\rho,1}=-\mathrm{diag}\{ \lambda_n \}_{n=N_0+1}^{N} -\rho I, \\
		{A}_{\gamma,1}= -\mathrm{diag}\{ \lambda_n \}_{n=N_0+1}^{N} -\gamma I,
		\end{array} \\
&		{\bf B}_1=\left[\begin{array}{cc}
		B_1  & 0 \\ 
		0  & B_1 
		\end{array}\right], ~~B_1=\mathrm{col}\{b_n\}_{n=N_0 +1}^{N}, \\ 
&		{\bf C}_1=\left[\begin{array}{cc}
		C_{z,1}  & 0 \\ 
		0  &  C_{v,1}
		\end{array}\right], ~ \begin{array}{cc}
		C_{z,1}=[c_{z,N_0+1}, \dots, c_{z,N}],\\
		C_{v,1}=[c_{v,N_0+1}, \dots, c_{v,N}],
		\end{array} \\ 
&	{\hat{X}}_1= \mathrm{col}\{\hat{z}_{N_0+1},\dots, \hat{z}_{N}, \hat{v}_{N_0+1},\dots,\hat{v}_{N}\}, \\
&	e_z^{N_0}=[e_{z,1}, \dots e_{z,N_0}]^{\mathrm{T}},~  		
		e_z^{N-N_0}=[e_{z,N_0+1}, \dots e_{z,N}]^{\mathrm{T}}, \\
&		e_v^{N_0}=[e_{v,1}, \dots, e_{v,N_0} ]^{\mathrm{T}},~ e_v^{N-N_0}=[ e_{v,N_0+1}, \dots, e_{v,N} ]^{\mathrm{T}} ,\\
&	e^{N_0}=\mathrm{col}\{ e_z^{N_0}, e_v^{N_0}\}, ~~ e^{N-N_0}=\mathrm{col}\{ e_z^{N-N_0}, e_v^{N-N_0}\}.	
	\end{split}
\end{equation}
From \eqref{error2}, we have  
\begin{equation}\label{eq225}
		E_{N-N_0}^{\varepsilon}\dot{e}^{N-N_0}(t)={\bf A}_1{e}^{N-N_0}(t).  
\end{equation}
By \eqref{observerdyeq44}, \eqref{observercontroller}, \eqref{observereq8}, and \eqref{error1}, we obtain the closed-loop system: 
\begin{subequations}\label{SPsec5eq1ss}
\begin{align}
&E_{N_0}^{\varepsilon}\dot{\hat{X}}_0(t)=({\bf A}_0-{\bf B}_0K_0) \hat{X}_0(t)  \nonumber \\
&~~~~~~~ +{\bf L}_0[{\bf C}_0e^{N_0}(t)+{\bf C}_1e^{N-N_0}(t)+\zeta(t)], \label{SPsec5eq1ss11}\\
&		E_{N_0}^{\varepsilon}\dot{e}^{N_0}(t)=({\bf A}_0-{\bf L}_0{\bf C}_0 )e^{N_0}(t)  \nonumber\\
&~~~~~~~ -{\bf L}_0{\bf C}_1e^{N-N_0}(t)-{\bf L}_0\zeta(t), \label{SPsec5eq1ss22}\\
&		E_1^{\varepsilon} \dot{X}_{n}(t) =A_{n}X_{n}(t) - B_{n}K_0\hat{X}_0(t), ~n>N,  \label{SPsec5eq1ss33}
\end{align}
\end{subequations} 
and \eqref{eq225} with $\zeta(t)$ subject to \eqref{zetazeta}. Moreover, ${\hat{X}}_1(t)$ satisfies  
\begin{equation}\label{eq224}
		E_{N-N_0}^{\varepsilon}\dot{\hat{X}}_1(t)={\bf A}_1 {\hat{X}}_1(t)-{\bf B}_1K_0 {\hat{X}}_0(t).  
\end{equation}
Later, we will prove that \eqref{eq224} is exponentially decaying provided $\hat{X}_0(t)$ is exponentially decaying through the Lyapunov method and S-procedure (see \eqref{eqX1}-\eqref{exstX1} below).
\begin{remark}
	For \eqref{eq224} with $\varepsilon>0$, since $E^{\frac{1}{\varepsilon}}_{N-N_0}{\bf A}_1$ is not necessarily Hurwitz, the exponential decay of ${\hat{X}}_1(t)$ does not follow directly from the exponential decay of $\hat{X}_0(t)$. 
	Hence, a dedicated Lyapunov analysis (see \eqref{eqX1}-\eqref{exstX1} below) is required, which is different from \cite{katz2022delayed} for heat equations. 
\end{remark}

For the stability of \eqref{eq225} and \eqref{SPsec5eq1ss}, we consider the Lyapunov function  
\begin{align}\label{SPLyapunovreduced}
	V(t)=&\,V_0(t) +V_1(t)+ V_{\infty}(t) ,   \\
	V_0(t) =&\, \hat{X}_0^{\mathrm{T}}(t){E}_{N_0}^{\varepsilon} P_{0}^{\varepsilon}\hat{X}_0(t) + (e^{N_0}(t))^{\mathrm{T}}{E}_{N_0}^{\varepsilon} P_{e}^{\varepsilon} e^{N_0}(t), \nonumber\\
	V_1(t) =&\,p({e}^{N-N_0}(t))^{\mathrm{T}}{E}_{N-N_0}^{\varepsilon} P_{1}^{\varepsilon}{e}^{N-N_0}(t),  \nonumber \\
	V_{\infty}(t)=&\,\sum_{n=N+1}^{\infty} X_{n}^{\mathrm{T}}(t)E_{1}^{\varepsilon}X_{n}(t),  \nonumber
\end{align}
where $p>0$ and matrices 
\begin{equation}\label{PPP}
	\begin{split}
	&	P_{0}^{\varepsilon} =\left[\begin{array}{cc}
		P_{0,1} & \varepsilon P_{0,2}^{\mathrm{T}}\\
		P_{0,2} & P_{0, 3}
		\end{array}\right],~ P_{0,1}>0,P_{0,3}>0,\\
	&P_{e}^{\varepsilon} = \left[\begin{array}{cc}
		P_{e,1} & \varepsilon P_{e,2}^{\mathrm{T}}\\
		P_{e,2} & P_{e,3}
		\end{array}\right],~~ P_{e,1}>0,~P_{e,3}>0, \\
	& P_{1}^{\varepsilon} =\left[\begin{array}{cc}
		P_{1,1} & \varepsilon P_{1,2}^{\mathrm{T}}\\
		P_{1,2} & P_{1,3}
		\end{array}\right],~~ P_{1,1}>0,~P_{1,3}>0,  
	\end{split}
\end{equation}
{\color{black}correspond to \eqref{SPsec5eq1ss11}, \eqref{SPsec5eq1ss22}, and \eqref{eq225}, respectively. We refer to $V_0$, $V_1(t)$, and $V_{\infty}$ as the finite-dimensional, intermediate, and infinite-dimensional parts of the Lyapunov function, respectively.}
Assume  
\begin{equation}\label{SPineq1}
		{E}_{N_0}^{\varepsilon} P_{0}^{\varepsilon}>0,{E}_{N_0}^{\varepsilon} P_{e}^{\varepsilon}>0, {E}_{N-N_0}^{\varepsilon}P_{1}^{\varepsilon}>0 ~\text{for} ~ \varepsilon>0.
\end{equation}
In \eqref{SPLyapunovreduced}, the construction of $V_0(t)$ is inspired by \cite{fridman2002effects,fridman2006robust} for SP ODEs. 
Our Lyapunov functions are consistent with the functions for the corresponding descriptor systems which allows to achieve efficient bounds on $\varepsilon$ and to prove feasibility of LMIs. 
Let $e_z^{N}=\mathrm{col}\{e_z^{N_0},e_z^{N-N_0}\}$, $e_v^{N}=\mathrm{col}\{e_v^{N_0},{\color{black}e_v^{N-N_0}}\}$. From \eqref{SPLyapunovreduced}, \eqref{SPineq1}, and Parseval's identity, we have for $\varepsilon\geq 0$, 
\begin{align}\label{sanjiao}
	V(t) & \geq  M_{1,s}\left[|\hat{z}^{N_0}(t)|^2+ |e_z^{N}(t)|^2 + \sum_{n=N+1}^{\infty}z_n^2(t) \right] \nonumber \\
	&+\varepsilon M_{1,f}\left[|\hat{v}^{N_0}(t)|^2+|e_v^{N}(t)|^2+  \sum_{n=N+1}^{\infty}v_n^2(t) \right], t\geq 0, \nonumber  \\
 V(0) & \leq M_{2,s}\|z_0\|^2_{L^2} + \varepsilon M_{2,f}\|v_0\|^2_{L^2}, 	
\end{align}
for some constants {\color{black}$0<M_{1,s}\leq M_{2,s}$ and $0<M_{1,f}\leq M_{2,f}$ which are independent of $\varepsilon$.} 
Differentiating $V(t)$ along solutions of \eqref{eq225} and \eqref{SPsec5eq1ss},  we have  
\begin{align}\label{SPineq2aabbcc}
	&\dot{V}(t) +2\delta V(t) = 
	\hat{X}_0^{\mathrm{T}}(t)\Psi^{(11)}_{\varepsilon}\hat{X}_0(t) \nonumber \\
	&+2 \hat{X}_0^{\mathrm{T}}(t) (P_{0}^{\varepsilon})^{\mathrm{T}}{\bf L}_0[{\bf C}_0e^{N_0}(t)+{\bf C}_1e^{N-N_0}(t)+\zeta(t)] \nonumber \\
	&- 2(e^{N_0}(t))^{\mathrm{T}} (P_{e}^{\varepsilon})^{\mathrm{T}} {\bf L}_0[{\bf C}_1e^{N-N_0}(t)+\zeta(t) ] \nonumber \\
	&+ (e^{N_0}(t))^{\mathrm{T}}\Psi^{(22)}_{\varepsilon}e^{N_0}(t) + (e^{N-N_0}(t))^{\mathrm{T}}\Psi^{(33)}_{\varepsilon}e^{N-N_0}(t) \nonumber \\
	&+  \sum_{n=N+1}^{\infty} X_n^{\mathrm T}(t)[A_n+A_n^{\mathrm T} +2{\delta}E_1^{\varepsilon} ]X_n(t) \nonumber \\
	&- \sum_{n=N+1}^{\infty}2X_n^{\mathrm T}(t)B_n K_0\hat{X}_0(t),  
\end{align}
where  
\begin{equation}\label{psipsi}
	\begin{split}
		\Psi^{(11)}_{\varepsilon} =&\, (P_{0}^{\varepsilon})^{\mathrm{T}} ({\bf A}_0-{\bf B}_0K_0)  +2\delta {E}_{N_0}^{\varepsilon} P_{0}^{\varepsilon} \\
		&\, +({\bf A}_0-{\bf B}_0K_0)^{\mathrm{T}}P_{0}^{\varepsilon},\\
		\Psi^{(22)}_{\varepsilon} =&\, (P_{e}^{\varepsilon})^{\mathrm{T}} ({\bf A}_0-{\bf L}_0{\bf C}_0) +2\delta {E}_{N_0}^{\varepsilon} P_{e}^{\varepsilon}\\
		&\, +({\bf A}_0-{\bf L}_0{\bf C}_0)^{\mathrm{T}}P_{e}^{\varepsilon},\\
		\Psi^{(33)}_{\varepsilon}=&\, p [(P_{1}^{\varepsilon})^{\mathrm{T}}{\bf A}_1+{\bf A}_1^{\mathrm{T}}P_{1}^{\varepsilon}  +2\delta {E}_{N-N_0}^{\varepsilon} P_{1}^{\varepsilon} ].  
	\end{split}
\end{equation}
By Young's inequality, we have for $0<R<2I_2$,  
\begin{equation}\label{observerdyeq2.15}
\begin{split}
		&\, -\sum_{n=N +1}^{\infty}2X_n^{\mathrm T}B_n K_0\hat{X}_0\\
		\leq &\, \sum_{n=N+1}^{\infty} \lambda_n X_n^{\mathrm T}RX_n  + \kappa_{N}\hat{X}_{0}^{\mathrm{T}}{K}_0^{\mathrm{T}} R^{-1} {K}_0\hat{X}_0,  
	\end{split} 
\end{equation}
where $\kappa_N=\sum_{n=N+1}^{\infty}\frac{b_n^2}{\lambda_n}=O(\frac{1}{N})$, $N\to\infty$.
Let $\eta_0(t)= \mathrm{col}\{\hat{X}_0(t), e^{N_0}(t),\zeta(t) \}$, $\eta(t)=\mathrm{col}\{\eta_0(t), {e}^{N-N_0}(t)\} $. 
Substituting \eqref{observerdyeq2.15} into \eqref{SPineq2aabbcc}, and using \eqref{zetazeta} and S-procedure, we obtain for $a_1, a_2>0$,  
\begin{equation}\label{Vdot}
	\begin{split}
	&\dot{V}(t) +2\delta V(t)   + a_1[ \|c_z\|^{2}_{N} \sum_{n=N+1}^{\infty} z^2_n(t) -|\zeta_z(t)|^{2}] \\
&~~~~~~~ + a_2[ \|c_v\|^{2}_{N} \sum_{n=N+1}^{\infty}v^2_n(t) -|\zeta_v(t)|^{2}]  \\
 &\leq \eta^{\mathrm{T}}(t)\Psi_\varepsilon \eta(t) +  \sum_{n=N+1}^{\infty} X_n^{\mathrm T}(t)\Theta_{n}^{\varepsilon}X_n(t)<0, ~t\ge 0,
	\end{split}
\end{equation}
provided  
\begin{subequations}\label{LMIaa11}
\begin{align}
&\Psi_{\varepsilon}=\setlength{\arraycolsep}{1pt} {\small 	\left[\begin{array}{ccc|cc}
		\hat{\Psi}^{(11)}_{\varepsilon}   &  (P_{0}^{\varepsilon})^{\mathrm{T}}{\bf L}_0{\bf C}_0  &  (P_{0}^{\varepsilon})^{\mathrm{T}}{\bf L}_0 & (P_{0}^{\varepsilon})^{\mathrm{T}}{\bf L}_0{\bf C}_1 \\ 
		* & \Psi^{(22)}_{\varepsilon} &  -(P_{e}^{\varepsilon})^{\mathrm{T}}{\bf L}_0 & -(P_{e}^{\varepsilon})^{\mathrm{T}}{\bf L}_0{\bf C}_1 \\
			* & *  & -\mathrm{diag}\{a_1, a_2\}& 0 \\ \hline
		*& *&*& \Psi^{(33)}_{\varepsilon} 
		\end{array}\right] }<0,  \label{LMIaa11a}\\
& \Theta_{n}^{\varepsilon}= A_n+A_n^{\mathrm T}+ 2\delta E_1^{\varepsilon}+  \lambda_n  R \nonumber \\
& ~~~~~~~ +  \mathrm{diag}\{a_1\|c_z\|^{2}_{N}, a_2\|c_v\|^{2}_{N}\} <0 , ~n>N,  \label{LMIaa11b}
\end{align}
\end{subequations} 
where $\hat{\Psi}^{(11)}_{\varepsilon}=\Psi^{(11)}_{\varepsilon}+ \kappa_{N}{K}_0^{\mathrm{T}}R^{-1}{K}_0$. 
Let $P_{1}^{\varepsilon}$, defined in \eqref{PPP}, satisfy  
\begin{equation}\label{P1epsP1}
	\begin{array}{ll}
		(P_{1}^{\varepsilon})^{\mathrm{T}}{\bf A}_1+{\bf A}_1^{\mathrm{T}}P_{1}^{\varepsilon}  +2\delta {E}_{N}^{\varepsilon} P_{1}^{\varepsilon} <0. 
	\end{array}
\end{equation}
By choosing $p>0$ large enough and using Schur complement, we find that \eqref{LMIaa11a} holds if and only if  
\begin{equation}\label{LMIcont}
	\begin{array}{ll}
	\tilde{\Psi}_\varepsilon:= \setlength{\arraycolsep}{1pt} { \left[ \begin{matrix}
	{\Psi}^{(11)}_{\varepsilon} &(P_{0}^{\varepsilon})^{\mathrm{T}}{\bf L}_0 {\bf C}_0 & (P_{0}^{\varepsilon})^{\mathrm{T}} {\bf L}_0 & K_0^{\mathrm{T}} \\
	*& \Psi^{(22)}_{\varepsilon} & -(P_{e}^{\varepsilon})^{\mathrm{T}} {\bf L}_0 & 0 \\
	* & *  & -\mathrm{diag}\{a_1 , a_2 \} & 0 \\
	*&*&*& - \frac{1}{\kappa_{N}} R
\end{matrix} \right]}<0.
	\end{array}
\end{equation}
Since $\{\lambda_n\}_{n=1}^{\infty}$ are monotonically increasing and $R\leq 2I$, \eqref{LMIaa11b} is feasible if and only if  
 \begin{equation}\label{LMItails}
 	\begin{split}
 \Theta_{N+1}^{\varepsilon}=&		A_{N+1}+A_{N+1}^{\mathrm T}+2\delta E_1^{\varepsilon} +  \lambda_{N+1}  R  \\
& +  \mathrm{diag}\{a_1\|c_z\|^{2}_{N}, a_2\|c_v\|^{2}_{N}\}<0.   
 	\end{split}
 \end{equation}
 Therefore, if \eqref{P1epsP1}, \eqref{LMIcont}, and \eqref{LMItails} are feasible for $\varepsilon\geq 0$, we can obtain from \eqref{Vdot} that 
 \begin{equation}\label{Vdot00}
 	V(t)\leq  \mathrm{e}^{-2\delta t}V(0), ~~t\ge 0.
 \end{equation}
 Next, we prove that for $\varepsilon\geq 0$, there exists $M_3(\varepsilon)\geq 1$ such that 
 \begin{equation}\label{X0expstable}
	\begin{split}
		&|\hat{X}_0(t)|^2+|e^{N}(t)|^2+\sum_{n=N+1}^{\infty}|X_n(t)|^2\\
		&\leq M_3(\varepsilon)\mathrm{e}^{-2\delta t} [\| z_0\|^2_{L^2}+ \varepsilon \|v_0\|^2_{L^2}],~t\ge 0.  
	\end{split}
\end{equation}

For $\varepsilon>0$, from \eqref{sanjiao} and \eqref{Vdot00}, it is easy to obtain \eqref{X0expstable} with $M_3(\varepsilon)=\frac{\max\{M_{2,s}, M_{2,f}\}}{\min\{M_{1,s}, \varepsilon M_{1,f}\}}\geq 1$. 
For $\varepsilon=0$, from \eqref{sanjiao} and \eqref{Vdot00} with $\varepsilon=0$, it follows
\begin{equation}\label{eq61}
	\begin{split}
	&|\hat{z}^{N_0}(t)|^2+ |e_z^{N}(t)|^2 + \sum_{n=N+1}^{\infty}z_n^2(t)\\
	&\leq M_{3,s}\mathrm{e}^{-2\delta t}\|z(\cdot,0)\|^2_{L^2}, ~~t\geq 0, 
	\end{split}
\end{equation}
where $M_{3,s}=\frac{M_{2,s}}{M_{1,s}}\geq 1$. This and \eqref{eq225} with $\varepsilon=0$ imply
\begin{equation}\label{evevev}
	|e_v^{N-N_0}(t)|^2\leq \bar{M}_{1}\mathrm{e}^{-2\delta t}\|z(\cdot,0)\|^2_{L^2},~ t\geq 0,
\end{equation}
for some $\bar{M}_{1}>0$.  
From \eqref{SPsec5eq1ss33} with $\varepsilon=0$ and $\lambda_n>-\gamma$, $n>N$ (see \eqref{fasteigenvalues}), we have $v_n(t)=  \frac{\beta}{\lambda_n+\gamma}z_n(t)-	\frac{b_n K_{21}\hat{z}^{N_0}(t)}{\lambda_n+\gamma}-\frac{b_n K_{22}\hat{v}^{N_0}(t)}{\lambda_n+\gamma}$. 
From the definition of $\zeta_v(t)$ in \eqref{observereq8}, it follows
\begin{equation}\label{zetavv}
	\zeta_v(t)= \sum_{n=N+1}^{\infty}\frac{\beta c_{v,n}}{\lambda_n+\gamma}z_n(t) -\bar{\kappa}_N [K_{21}\hat{z}^{N_0}(t)+ K_{22}\hat{v}^{N_0}(t)], 
\end{equation}
where $\bar{\kappa}_N=\sum_{n=N+1}^{\infty}\frac{c_{v,n}b_n}{\lambda_n+\gamma}$. From \eqref{SPsec5eq1ss11}, \eqref{SPsec5eq1ss22} with $\varepsilon=0$, and \eqref{zetavv}, we see that 
\begin{equation*}
\begin{split}
	& \left( \bar{A} +\bar{\kappa}_N { \left[ \begin{matrix}
	 - {\bf L}_{0}^{(22)}  \\
	   {\bf L}_{0}^{(22)} 
	 \end{matrix} \right]}[K_{22}, ~0] \right) {\small \left[ \begin{matrix}
	 	\hat{v}^{N_0}(t) \\
	 	e_{v}^{N_0}(t)
	 \end{matrix} \right]},\\
	&\bar{A} =	{	\left[ \begin{matrix}
	 	A_{\gamma,0}-B_0K_{22}  & {\bf L}_0^{(22)}C_{v,0} \\
	   0  & A_{\gamma,0}-{\bf L}_0^{(22)}C_{v,0} 
	 \end{matrix} \right]}, 
\end{split}
\end{equation*}
can be expressed as a linear combination of $\hat{z}^{N_0}(t)$, $e_z^{N}(t)$, $e_v^{N-N_0}(t)$, $\zeta_z(t)$ and $\sum_{n=N+1}^{\infty}\frac{\beta c_{v,n}}{\lambda_n+\gamma}z_n(t)$.
These terms decay exponentially with rate $\delta$ according to \eqref{eq61} and \eqref{evevev}. 
Then, if 
\begin{equation}\label{invitable}
\bar{A} +\bar{\kappa}_N {\small \left[ \begin{matrix}
	 - {\bf L}_{0}^{(22)}  \\
	   {\bf L}_{0}^{(22)} 
	 \end{matrix} \right]}[K_{22}, ~0] ~\text{is invertible},
\end{equation}
we arrive at
\begin{equation}\label{evevev22}
	|\hat{v}^{N_0}(t)|^2+|e_{v}^{N_0}(t)|^2\leq \bar{M}_{2}\mathrm{e}^{-2\delta t}\|z(\cdot,0)\|^2_{L^2},~ t\geq 0,
\end{equation}
for some $\bar{M}_{2}>0$. Note that \eqref{invitable} always holds for sufficiently large $N$ since $\bar{A}$ is invertible and $\bar{\kappa}_N\to 0$, $N\to \infty$. From \eqref{eq61}, \eqref{evevev}, \eqref{evevev22}, and the expression for $v_n$ below \eqref{evevev}, we get \eqref{X0expstable} for some constant $M_{3}(0)\ge 1$.

For the stability of \eqref{eq224}, we consider  
\begin{equation}\label{eqX1}
		V_{X_1}(t)= \hat{X}^{\mathrm{T}}_1(t)  {E}_{N-N_0}^{\varepsilon} P_{1}^{\varepsilon} \hat{X}_1(t),  
\end{equation}
where $P_{1}^{\varepsilon}$ is defined in \eqref{PPP} and satisfies \eqref{P1epsP1}. 
The derivative of $V_{X_1}(t)$ along the solutions of \eqref{eq224} yields  
\begin{equation}\label{deriveVX1}
	\begin{split}
		&\dot{V}_{X_1}(t)+2\delta V_{X_1}(t)- \chi | {\bf B}_1 K_0\hat{X}_0(t)|^2  \\
		&= \left[\begin{array}{cc}
		\hat{X}_1(t) \\
		{\bf B}_1K_0\hat{X}_0(t)
		\end{array}\right]^{\mathrm{T}} \Psi_1  \left[\begin{array}{cc}
		\hat{X}_1(t) \\
		{\bf B}_1K_0\hat{X}_0(t)
		\end{array}\right]<0,  
	\end{split}
\end{equation}
if  
\begin{equation*}
	\Psi_1	=\left[\begin{array}{cc}
		(P_{1}^{\varepsilon})^{\mathrm{T}}{\bf A}_1+{\bf A}_1^{\mathrm{T}}P_{1}^{\varepsilon}  +2\delta {E}_{N}^{\varepsilon} P_{1}^{\varepsilon} & -(P_{1}^{\varepsilon})^{\mathrm{T}} \\
		*& -\chi I
		\end{array}\right]<0.   
\end{equation*}
Since $P_{1}^{\varepsilon}$ satisfies \eqref{P1epsP1}, we can always fix a large enough $\chi>0$ such that $\Psi_1<0$. Moreover, \eqref{P1epsP1} holds with $\delta$ replaced by $\delta+  \delta_0$ if $\delta_0>0$ is small enough, i.e., \eqref{deriveVX1} holds with $\delta$ replaced by $\delta+  \delta_0$. Thus,  
\begin{equation}\label{eq2.39}
	\begin{split}
		\dot{V}_{X_1}(t) \leq&\, -2(\delta +\delta_0 ){V}_{X_1}(t)+\chi| {\bf B}_1 K_0\hat{X}_0(t)|^2\\
		\overset{\eqref{X0expstable}}{\leq}&\, M_4(\varepsilon) \mathrm{e}^{-2\delta t} [\| z(\cdot,0)\|^2_{L^2}+ \|v(\cdot,0)\|^2_{L^2}]  \\
	 &\, -2(\delta +\delta_0 ){V}_{X_1}(t) , ~~t\geq 0,  
	\end{split}
\end{equation}
where $M_4(\varepsilon)=\chi|{\bf B}_1 K_0|^2 M_3(\varepsilon)$. 
 The comparison principle implies that  
\begin{equation}\label{X1expstable}
	\begin{split}
		{V}_{X_1}(t) \leq&\, \mathrm{e}^{-2\delta t}{V}_{X_1}(0) \\
		&\, + \frac{M_4(\varepsilon)}{2\delta_0} \mathrm{e}^{-2\delta t} [\| z_0\|^2_{L^2}+ \|v_0\|^2_{L^2}] , ~t>0.    
	\end{split}
\end{equation}
From \eqref{eqX1} and \eqref{X1expstable}, it follows  
\begin{equation}\label{exstX1}
		|\hat{X}_{1}(t)|^2 \leq M_{5}(\varepsilon)\mathrm{e}^{-2\delta t}[\| z_0\|^2_{L^2}+ \varepsilon \|v_0\|^2_{L^2}], ~t\geq 0,
\end{equation}
for  some $M_5(\varepsilon)>0$. Combining \eqref{X0expstable}, \eqref{exstX1}, and  Parseval's identity, we arrive at  
\begin{equation}\label{expstable}
	\begin{split}
		&\|z(\cdot,t)\|^{2}_{L^2}+\|z(\cdot,t)-\hat{z}(\cdot,t)\|^{2}_{L^2}\\
		&+\|v(\cdot,t)\|^{2}_{L^2}+\|v(\cdot,t)-\hat{v}(\cdot,t)\|^{2}_{L^2}\\
		\leq&\, M(\varepsilon) \mathrm{e}^{-2\delta t}[ \|z_0\|^{2}_{L^2}+\varepsilon \|v_0\|^{2}_{L^2} ], ~~t>0,  
	\end{split}
\end{equation}
for $\varepsilon\geq 0$ and some $M(\varepsilon)\geq 1$.

\subsection{The LMI design of $K_0$ from slow PDE system \eqref{eq1slow}  }\label{sec2.5}
Consider inequalities \eqref{PPP}, \eqref{P1epsP1}, \eqref{LMIcont}, and \eqref{LMItails} with $\varepsilon=0$.  
First, from the definition of $N_0$ in \eqref{NNN}, we can obtain that $(E_{N-N_0},{\bf A}_1+{\delta}I)$ is stable. Hence, there always exists $P_{1}^{0}$ such that \eqref{P1epsP1} with $\varepsilon=0$ is feasible (see \cite[Theorem 2.1]{xu2006robust}). 
For $\tilde{\Psi}_0<0$ in \eqref{LMIcont}, we introduce notations $Q_0=(P_{0}^{0})^{-1}$ and $Y=K_0Q_0$. 
Considering the structures of $Q_0$ and $P_{e}^{0}$, we can represent them as  
\begin{equation*}
	\begin{split}
		&Q_0={\bf P}_0{E}_{N_0}+S{\bf Q}_0, ~S=[ 0_{N_0 \times N_0} ,  I_{N_0} ]^{\mathrm{T}},\\
		&P_{e}^0={\bf P}_e{E}_{N_0}+S{\bf Q}_e,  
	\end{split}
\end{equation*}
where $0<{\bf P}_0,{\bf P}_e\in\mathbb{R}^{2N_0 \times 2N_0}$, ${\bf Q}_0=[{\bf Q}_{01}, {\bf Q}_{02} ]$, ${\bf Q}_e=[{\bf Q}_{e1}, {\bf Q}_{e2} ]$ with ${\bf Q}_{01}, {\bf Q}_{e1}  \in \mathbb{R}^{N_0\times N_0}$ and $0<{\bf Q}_{02}, {\bf Q}_{e2} \in \mathbb{R}^{N_0 \times N_0}$.
Multiplying $\tilde{\Psi}_0$ defined in \eqref{LMIcont}  by $\mathrm{diag}\{Q_0,I,I,I\}$ and its transpose, from the right and the left, 
we find that $\tilde{\Psi}_0<0$ if and only if  
\begin{equation}\label{dyLMIa100ss}
	 \begin{split}
	&	\setlength{\arraycolsep}{2pt}{ \left[ \begin{matrix}
	\tilde{\Psi}^{(11)} &{\bf L}_0 {\bf C}_0 &  {\bf L}_0 &  Y^{\mathrm{T}} \\
	*& \tilde{\Psi}^{(22)} & -({\bf P}_e{E}_{N_0}+S{\bf Q}_e)^{\mathrm{T}} {\bf L}_0 & 0 \\
	* & *  & -\mathrm{diag}\{a_1, a_2\} & 0 \\
	*&*&*& -\frac{1}{\kappa_N}R
\end{matrix} \right]}<0 ,\\	
&	\tilde{\Psi}^{(11)}=	{\bf A}_0^{\delta} ({\bf P}_0{E}_{N_0}+S{\bf Q}_0)-{\bf B}_0Y   \\
&	~~~~~~ +({\bf P}_0{E}_{N_0}+S{\bf Q}_0)^{\mathrm{T}}({\bf A}_0^{\delta})^{\mathrm{T}}  -Y^{\mathrm{T}}{\bf B}_0^{\mathrm{T}},\\
&	\tilde{\Psi}^{(22)}=	 ({\bf P}_e{E}_{N_0}+S{\bf Q}_e)^{\mathrm{T}}({\bf A}_0^{\delta}-{\bf L}_0{\bf C}_0)\\
&	~~~~~~  +({\bf A}_0^{\delta}-{\bf L}_0{\bf C}_0)^{\mathrm{T}}({\bf P}_e{E}_{N_0}+S{\bf Q}_e).  
\end{split}
\end{equation} 
If LMIs \eqref{LMItails} with $\varepsilon=0$ and \eqref{dyLMIa100ss}, with decision variables ${\bf P}_0$, ${\bf Q}_{01}$, ${\bf Q}_{02}$, ${Y}$, ${\bf P}_e$, ${\bf Q}_{e1}$, ${\bf Q}_{e2}$, $R$, $a_1, a_2$, are feasible, the controller gain is obtained by  
\begin{equation}\label{KKss}
		K_0=\left[ \begin{matrix}
	K_{11} & K_{12}  \\
	K_{21}& K_{22}
\end{matrix} \right] =Y({\bf P}_0{E}_{N_0}+S{\bf Q}_0)^{-1}. 
\end{equation} 
 
Moreover, if \eqref{LMItails} with $\varepsilon=0$ and \eqref{dyLMIa100ss} are feasible with $K_{22}$ given in \eqref{KKss}, the following fast LMIs  are feasible:  
 \begin{align}\label{fasterLMIcont}
	& \left[ \begin{matrix}
	{\Psi}^{(11)}_f &P_{0,3}{\bf L}_0^{(22)} {\bf C}_v & P_{0,3}^{\mathrm{T}} {\bf L}_0^{(22)} & K_{22}^{\mathrm{T}} \\
	*& \Psi_{f}^{(22)} & -P_{e,3}{\bf L}_0^{(22)} & 0 \\
	* & *  & - a_2  & 0 \\
	*&*&*& - \frac{1}{\kappa_{N}} r_{22}
\end{matrix} \right] <0, \nonumber \\
&-2\lambda_{N+1}-2\gamma+\lambda_{N+1}r_{22}+a_2\|c_{v}\|^{2}_{N}<0,  
\end{align}
 where  
 \begin{equation*}
 	\begin{split}
 		{\Psi}_{f}^{(11)}=P_{0,3}({\bf A}_{\gamma}-B_0K_{22})+({\bf A}_{\gamma}-BK_{22})^{\mathrm{T}}P_{0,3},\\
 		{\Psi}_{f}^{(22)}=P_{e,3}({\bf A}_{\gamma}-{\mathcal L}_{22}C_v)+({\bf A}_{\gamma}-{\mathcal L}_{22}C_v)^{\mathrm{T}}P_{e,3}.  
 	\end{split}
 \end{equation*}
 The feasibility of \eqref{fasterLMIcont} with $0<P_{0,3}, P_{e,3}\in\mathbb{R}^{N_0\times N_0}$, $a_2,r_{22}>0$ implies that the fast system \eqref{eq1reducedfaster} with observer  
 \begin{equation}\label{fasterobserver}
 	\begin{split}
 	&	\hat{v}(x,t)=\sum_{n=1}^{N}\hat{v}_{n}(t)\phi_{n}(x),\\
 	&	\dot{\hat{v}}_n(t) = (-{\lambda}_n-\gamma) \hat{v}_n(t)  +  b_{n}  {u}_2(t),  
 	\end{split}
 \end{equation}
 and controller   
  \begin{equation}\label{fastercontroller}
 	\begin{array}{ll}
 	{ u}_2(t)=-K_{22}\hat{v}^{N_0}(t), ~\hat{v}^{N}=\mathrm{col}\{\hat{v}_1,\dots,\hat{v}_{N_0}\},  
 	\end{array}
 \end{equation}
 is exponentially stable with a small enough decay rate. 
  
 \subsection{Main results}\label{II-F}
{\color{black} Let $\varepsilon_0=0$, $\varepsilon_1>0$, and
 \begin{equation*}
 	P_{\mu}^{\varepsilon_i,i} = \left[\begin{array}{cc}
		P_{\mu,1}^{(i)} & \varepsilon_i P_{\mu,2}^{\mathrm{T}}\\
		P_{\mu,2} & P_{\mu,3}
		\end{array}\right], ~\mu\in \{0,e,1\}, i=0,1.
 \end{equation*}
Since $E_{N_0}^\varepsilon P_0^\varepsilon$, $E_{N_0}^\varepsilon P_e^\varepsilon$, and $E_{N-N_0}^\varepsilon P_1^\varepsilon$ are affine in $\varepsilon$, the positivity conditions \eqref{SPineq1} for $\varepsilon\in(0,\varepsilon_1]$
are guaranteed by
\begin{equation}\label{positive}
E_{N_0}^{\varepsilon_1}P_0^{\varepsilon_1,1}>0,\quad
E_{N_0}^{\varepsilon_1}P_e^{\varepsilon_1,1}>0,\quad
E_{N-N_0}^{\varepsilon_1}P_1^{\varepsilon_1,1}>0.
\end{equation}
 Moreover, since the matrices in \eqref{P1epsP1},
\eqref{LMIcont}, and \eqref{LMItails} are affine in $\varepsilon$,
their negativity over $\varepsilon\in[0,\varepsilon_1]$ is guaranteed
by checking the two endpoints $\varepsilon_i$, $i=0,1$.
Therefore, it is sufficient that
   \begin{subequations}\label{SPineq1abc}
\begin{align}
&(P_{1}^{\varepsilon_i,i})^{\mathrm{T}}{\bf A}_1+{\bf A}_1^{\mathrm{T}}P_{1}^{\varepsilon_i,i}  +2\delta {E}_{N}^{\varepsilon_i} P_{1}^{\varepsilon_i,i} <0,  \label{SPineq1111a}\\
& \left[ \begin{matrix}
	{\Psi}^{(11)}_{\varepsilon_i,i} &(P_{0}^{\varepsilon_i,i})^{\mathrm{T}}{\bf L}_0 {\bf C}_0 & (P_{0}^{\varepsilon_i,i})^{\mathrm{T}} {\bf L}_0 & K_0^{\mathrm{T}} \\
	*& \Psi^{(22)}_{\varepsilon_i,i} & -(P_{e}^{\varepsilon_i,i})^{\mathrm{T}} {\bf L}_0 & 0 \\
	* & *  & -\mathrm{diag}\{a_1 , a_2 \} & 0 \\
	*&*&*& - \frac{1}{\kappa_{N}} R
\end{matrix} \right] <0, \label{SPineq1111b}\\
&A_{N+1}+A_{N+1}^{\mathrm T}+ 2\delta E_{1}^{\varepsilon_i} + \lambda_{N+1}  R  \nonumber\\
& ~~~~~~~~~ +  \mathrm{diag}\{a_1\|c_z\|^{2}_{N}, a_2\|c_v\|^{2}_{N}\}<0, \label{SPineq1111c}
\end{align}
\end{subequations}   
  where 
  \begin{equation*}
 \begin{split}
&	{\Psi}^{(11)}_{\varepsilon_i,i}=	(P_{0}^{\varepsilon_i,i})^{\mathrm{T}}({\bf A}_0-{\bf B}_0K_0)+({\bf A}_0-{\bf B}_0K_0)^{\mathrm{T}}P_{0}^{\varepsilon_i,i}\\
&	~~~~~~~~  +2\delta {E}_{N_0}^{\varepsilon_i} P_{0}^{\varepsilon_i,i}, \\
&	\Psi^{(22)}_{\varepsilon_i,i}  = (P_{e}^{\varepsilon_i,i})^{\mathrm{T}} ({\bf A}_0-{\bf L}_0{\bf C}_0)+({\bf A}_0-{\bf L}_0{\bf C}_0)^{\mathrm{T}}P_{e}^{\varepsilon_i,i} \\
&	~~~~~~~~ +2\delta {E}_{N_0}^{\varepsilon_i} P_{e}^{\varepsilon_i,i}.  
	\end{split}
\end{equation*}
Note that \eqref{SPineq1111a} guarantees the exponential stability of the intermediate modes \eqref{eq225} and \eqref{eq224}, whereas \eqref{SPineq1111b} and \eqref{SPineq1111c} ensure the negative definiteness of the finite-dimensional and infinite-dimensional parts of the Lyapunov derivative, respectively. }
The feasibility of \eqref{positive} and \eqref{SPineq1abc} guarantees the exponential stability of \eqref{eq225}, \eqref{SPsec5eq1ss}, and \eqref{eq224} with decay rate $\delta$ for all $\varepsilon\in[0,\varepsilon_1]$.
Summarizing, we have proved the following results.
\begin{theorem}\label{thm4.1}
	Consider system \eqref{eq1} with measurement \eqref{measurement}, where $c_z,c_v \in L^2(0,1)$ satisfy \eqref{cneq0}, control law \eqref{observercontroller}, and initial value $z_0,v_0\in H^1(0,1)$.  Let $\delta>0$ be a desired decay rate, $N_0\in\mathbb{N}$ satisfy \eqref{NNN}, and $N\ge N_0$.  Let  ${\bf L}_0$ be found from \eqref{observergain11} and given by \eqref{L0L0}.  Then the following hold: 
	
	{\bf 1)} Let there exist constants $a_1,a_2>0$, matrices $0<R\in \mathbb{R}^{2\times 2}$, $0<{\bf P}_0,{\bf P}_e\in\mathbb{R}^{2N_0\times 2N_0}$, ${\bf Q}_0=[{\bf Q}_{01}, {\bf Q}_{02} ]$, ${\bf Q}_e=[{\bf Q}_{e1}, {\bf Q}_{e2} ]$ with ${\bf Q}_{01}, {\bf Q}_{e1}  \in \mathbb{R}^{N_0\times N_0}$ and $0<{\bf Q}_{02}, {\bf Q}_{e2} \in \mathbb{R}^{N_0 \times N_0}$ such that LMIs \eqref{LMItails} with $\varepsilon=0$ and \eqref{dyLMIa100ss} are feasible. Let $K_0$ be given by \eqref{KKss} such that \eqref{invitable} holds. 
  Then the descriptor PDE system \eqref{eq1slow} under the controller \eqref{observercontroller} and observer \eqref{finobserver}, \eqref{observerdyeq44} with $\varepsilon=0$ is exponentially stable with decay rate $\delta>0$, and the fast system \eqref{eq1reducedfaster} under observer \eqref{fasterobserver} and control law \eqref{fastercontroller} is exponentially stable with a small enough decay rate. 
  
  {\bf 2)}
Let the gain $K_0$ be given as in {\bf 1)}. Given
$0=\varepsilon_0<\varepsilon_1$, assume that there exist constants
$a_1,a_2>0$, matrices
$0<P_{0,1}^{(i)},P_{0,3},P_{e,1}^{(i)},P_{e,3}
\in\mathbb{R}^{N_0\times N_0}$,
$0<P_{1,1}^{(i)},P_{1,3}
\in\mathbb{R}^{(N-N_0)\times(N-N_0)}$, $i=0,1$,
$P_{0,2},P_{e,2}\in\mathbb{R}^{N_0\times N_0}$, $P_{1,2}\in\mathbb{R}^{(N-N_0)\times(N-N_0)}$, and $0<R\leq2I$ such that \eqref{positive} and LMIs \eqref{SPineq1abc} are feasible. Then, for every $\varepsilon\in(0,\varepsilon_1]$, the solutions $z(x,t)$ and $v(x,t)$ of \eqref{eq1} under the control law \eqref{observercontroller}, together with the corresponding observer states $\hat z(x,t)$ and $\hat v(x,t)$ defined by \eqref{finobserver}, satisfy \eqref{expstable} with some $M(\varepsilon)\geq1$. 
\end{theorem}

 Theorem \ref{thm4.1} provides constructive LMI conditions for finding the observer dimension $N$ and maximal singular perturbation parameter $\varepsilon_1$ that guarantee stability.
We next state that the LMI conditions \eqref{SPineq1abc} always admit a feasible solution with the proof given in Appendix \ref{proofProposition}.  
\begin{proposition}\label{propos2.1}
The LMI conditions \eqref{SPineq1abc} are always feasible for small enough $\varepsilon_1>0$ and large enough $N\in \mathbb{N}$.
\end{proposition}

\begin{remark}\label{remark5}
For infinite-dimensional systems, exponential stability of the slow and fast subsystems may not guarantee stability of the full SP system. In \cite{tang2017stability}, a counterexample was given for a hyperbolic PDE coupled with a fast ODE, where the transport equation introduces an exact propagation delay which, together with the boundary coupling, leads to an unstable difference equation.  The situation is different from the SP parabolic systems studied in this paper. To illustrate this, consider the state-feedback case presented in \cite{pengfei2025CDC}. After modal decomposition, the dynamics of the first $N$ modes form a $2N$-dimensional SP ODE system, for which stability of the slow and fast subsystems guarantees stability for sufficiently small $\varepsilon>0$ by classical singular perturbation results for ODEs. Moreover, the residual modes are increasingly dissipative as $\lambda_n\to\infty$, and their convergence is guaranteed by the Lyapunov analysis for sufficiently small $\varepsilon>0$. 
\end{remark}

\section{Sampled-data control under discrete-time measurements}\label{III}
{\color{black} In this section, we present sampled-data implementation of  the continuous-time controller developed in Section~\ref{sec2}. The controller and observer gains obtained in Section~II are retained, while the continuous-time control inputs and measurements are replaced by their sampled-data counterparts. This allows us to investigate how the sampling of the slow and fast variables affects the stability conditions established in the continuous-time case.}

\begin{figure*}[!htb]
\centering
\includegraphics[width=10.8cm]{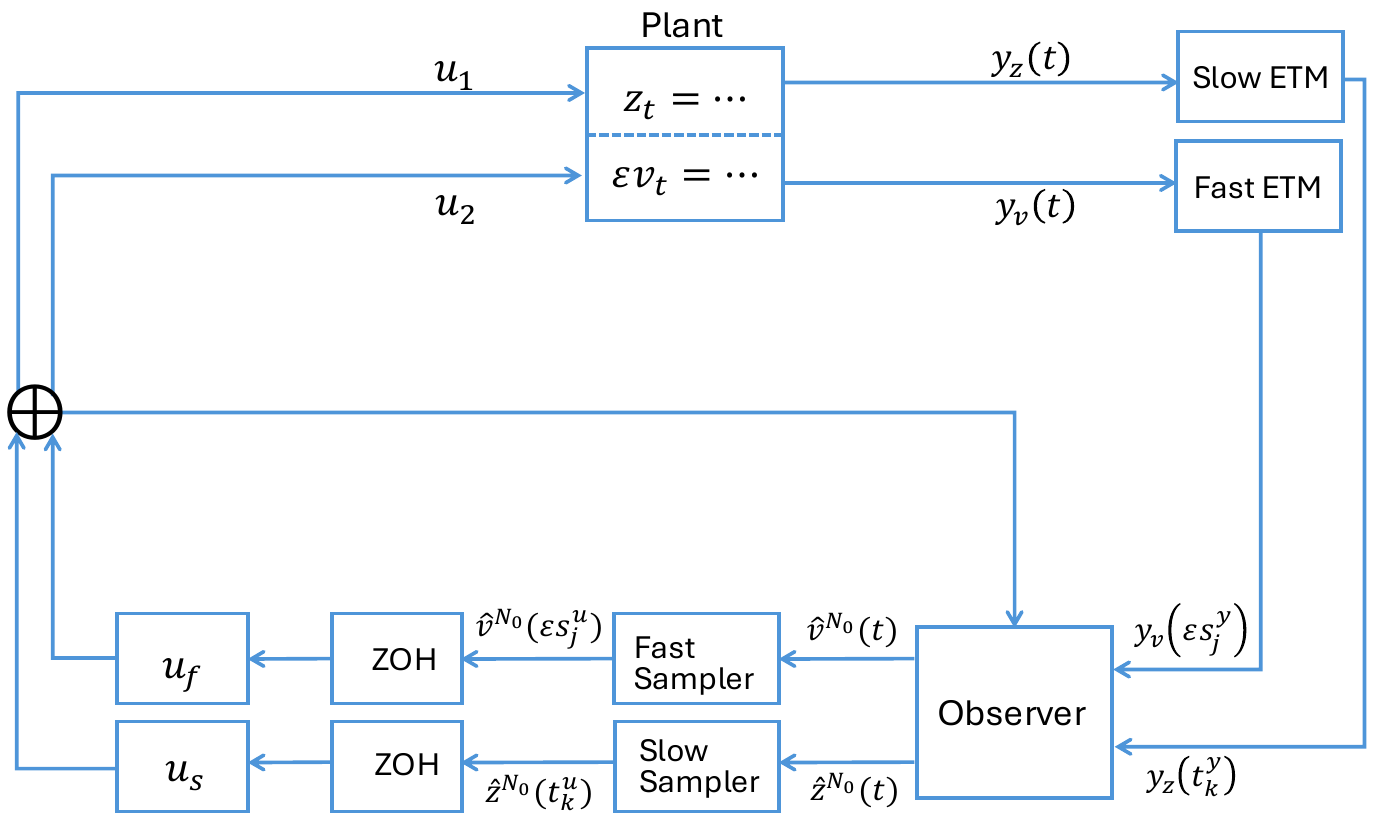} 
\caption{Block diagram representing the sampled-data control with event-triggered measurement.}\label{fig0}
\end{figure*} 
Given $\varepsilon>0$, we will use slow sampling instants $t_k^y$ (for the measurements) and $t_k^u$ (for the control) and fast sampling instants $\varepsilon s_j^y$  for $y$ and  $\varepsilon s_j^u$  for $u$ (see Fig. \ref{fig0}). It is physically meaningful to use slow and fast sampling as this matches the natural time-scale separation of the system. This multi-rate sampled-data framework is more efficient and realistic for digital implementation. It was reported in \cite{bidani2002multirate} that the use of multi-rate controllers can provide better performance than single-rate controllers. We will employ event-triggered sampling for $y$ and time-triggered sampling for $u$. 

Consider slow sampling instants $t_k^y$ and fast sampling instants $\varepsilon s_j^y$ (which are to be determined later from ETM) corresponding to the measurements:
\begin{equation*}
	\begin{array}{ll}
		0=t_0^y<t_1^y<\dots <t_k^y< t_{k+1}^y<\dots,\\
		0=\varepsilon s_0^y<\varepsilon s_1^y< \dots <\varepsilon s_j^y <\varepsilon s_{j+1}^y < \dots.
	\end{array}
\end{equation*} 
We consider system \eqref{eq1} (i.e., system \eqref{studiedsystem2.4}) with the measurement transmitted to the observer via a digital communication channel, given by 
\begin{equation}\label{measurementSD}
	\begin{split}
	&y(t)= [y_z(t_k^{y}), y_v(\varepsilon s^y_j)]^{\mathrm{T}}, ~ t\in [t_k^y, t_{k+1}^{y}) \cap [\varepsilon s_j^y, \varepsilon s_{j+1}^{y}),\\	   
	&y_z(t)=\int_{0}^{1}c_z(x) z(x,t)\mathrm{d}x, ~y_v(t)=\int_{0}^{1}c_v(x) v(x,t)\mathrm{d}x, 
	\end{split}
\end{equation}
where $c_z, c_v \in L^2(0,1)$.

Let $\delta>0$ be a desired decay rate and $N_0\in\mathbb{N}$ satisfy \eqref{NNN}. 
We consider the finite-dimensional observer of the form \eqref{finobserver} with $N\geq N_0$, where $\hat{X}_n=[\hat{z}_n(t), \hat{v}_n(t)]^\mathrm{T}$ ($1\leq n\leq N$) satisfy  
\begin{align}\label{observerdyeq44SD}
	&E^\varepsilon_1 \dot{\hat{X}}_{n}(t) =A_n \hat{X}_{n}(t) +  b_{n}{u}(t) - L_n  \left[ \begin{array}{cc}
		\hat{y}_z(t) -y_z(t_k^y)\\
		\hat{y}_v(t)-y_v(\varepsilon s_j^y)
		\end{array} \right] ,  \nonumber \\
	&\quad\quad\quad\quad\quad\quad\quad\quad\quad\quad\quad\quad  t\in [t_k^y, t_{k+1}^{y}) \cap [\varepsilon s_j^y, \varepsilon s_{j+1}^{y}), \nonumber \\	
	&\hat{X}_{n}(0)=0.   
\end{align}
Here $\hat{y}_z(t)=\int_{0}^{1}c_z(x) \hat{z}(x,t)\mathrm{d}x$,  $\hat{y}_v(t)=\int_{0}^{1}c_v(x) \hat{v}(x,t)\mathrm{d}x$, $y_z(t)$ and $y_v(t)$ are defined in \eqref{measurementSD}, $L_{n}\in\mathbb{R}^{2\times 2}$, $n=1,\dots, N_0$, are the observer gains obtained from \eqref{observergain11} and given by \eqref{L0L0}. Let $L_n=0_{2\times 2}$ for $n=N_0+1,\dots, N$.

Denote the error ${\bf d}_{\varepsilon}(t)=[d_z(t),d_{v,\varepsilon}(t)]^{\mathrm{T}}$ by   
\begin{equation}\label{dzdv}
	\begin{array}{ll}
	d_z(t) = y_z(t) - y_z(t_k^y) , ~~t\in 	[t_k^y, ~ t_{k+1}^{y}),\\
	d_{v,\varepsilon}(t) = y_v(t) - y_v(\varepsilon s_j^{y}) , ~~t\in[\varepsilon s_j^y,\varepsilon s_{j+1}^y).
	\end{array}
\end{equation}
Then the last term in \eqref{observerdyeq44SD} can be represented as  
\begin{equation}\label{observereq8SD}
	\begin{split}
		& \left[ \begin{array}{cc}
		\hat{y}_z(t) -y_z(t_k^y)\\
		\hat{y}_v(t)-y_v(\varepsilon s_j^y)
		\end{array} \right] \\
		=& \left[ \begin{array}{cc}
		\int_{0}^{1}c_z(x)[\hat{z}(x,t)-z(x,t)] \mathrm{d}x + d_z(t)\\
		\int_{0}^{1}c_v(x)[\hat{v}(x,t)-v(x,t)] \mathrm{d}x+d_{v,\varepsilon}(t)
		\end{array} \right]\\
		=&-{\bf C}_0e^{N_0}(t)-{\bf C}_1e^{N-N_0}(t)-\zeta(t) + {\bf d}_{\varepsilon}(t), \\
	&\quad\quad\quad\quad\quad\quad	t\in [t_k^y, t_{k+1}^{y}) \cap [\varepsilon s_j^y, \varepsilon s_{j+1}^{y}),
	\end{split}
\end{equation}
where ${\bf C}_0$ is defined in \eqref{notationsN0}, ${\bf C}_1$, $e^{N_0}(t)$, and $e^{N-N_0}(t)$ are defined in \eqref{notation222}, and $\zeta(t)$ is defined in \eqref{observereq8}. 

\begin{remark}\label{remark3.1}
	Note that ${\bf d}_{\varepsilon}(t)$ in \eqref{observereq8SD} involves 
	$\zeta_{z}(t_k^y)=\sum_{n=N+1}^{\infty}c_{z,n}z_n(t_k^y)$ and $\zeta_{v}(\varepsilon s_j^y)=\sum_{n=N+1}^{\infty}c_{v,n}v_n(\varepsilon s_j^y)$. In the existing literature \cite{katz2022delayed,katzasampled} that employs the time-delay approach to sampled-data control and Lyapunov functions with
	$V_{\mathrm{tail}}(t)=\sum_{n=N+1}^{\infty}|z_n(t)|^2+ \varepsilon \sum_{n=N+1}^{\infty}|v_n(t)|^2$, such terms are compensated by using Halanay's inequality, where the following bounds are employed:  
	\begin{align*}
		&-2\delta_1 \sup_{ t-\tau_z  \leq s \leq t} V_{\mathrm{tail}}(s ) \leq -2\delta_1 V_{\mathrm{tail}}(t_k^y) \\
		\leq& -2\delta_1\sum_{n=N+1}^{\infty} z_{n}^{2}(t_k^y) \leq -\frac{2\delta_1}{\|c_z\|^2} \zeta_z^2(t_k^y), ~~t\in [t_k^y,t_{k+1}^y), \\
	&-2\delta_2 \sup_{ t- \varepsilon \tau_v \leq s \leq t} V_{\mathrm{tail}}(s) \leq -2\delta_2 V_{\text{tail}}(\varepsilon s_j^y) \\
		\leq& -2\delta_2\sum_{n=N+1}^{\infty} \varepsilon v_{n}^{2}(\varepsilon s_j^y) \leq -\frac{2\delta_2 \varepsilon }{\|c_v\|^2} \zeta_v^2(\varepsilon s_j^y), ~~t\in [\varepsilon s_j^y,\varepsilon s_{j+1}^y).  
	\end{align*}
However, for sufficiently small $\varepsilon>0$, $-\frac{2\delta_2 \varepsilon}{\|c_v\|^2}$ cannot compensate the term
$\zeta_v(\varepsilon s_j^y)$ coupled with the finite-dimensional state of the closed-loop system. As a result, the Halanay-based approach employed in the previous works is no longer applicable. 
This motivates the event-triggered measurement strategy suggested in this paper. 
\end{remark}

For producing the measurement sampling instants, we employ the following ETMs (see Fig. \ref{fig0}):  
\begin{subequations}\label{ETmechanism}
\begin{align}
&t_{k+1}^y=\min\{t> t_k^y|   d_z^2(t) \geq \omega_1 (|y_z(t)-\hat{y}_z(t)|^2+\mathrm{e}^{-2\theta t})\}, \label{ETmechanismslow} \\
&\varepsilon s_{j+1}^y=\min\{t> \varepsilon s_j^y|  d_{v,\varepsilon}^2(t) \geq  \omega_2 (|y_v(t)-\hat{y}_v(t)|^2  \nonumber \\
& \quad\quad\quad\quad\quad + \mathrm{e}^{-2\theta t}) \}, \label{ETmechanismfast}
\end{align}
\end{subequations} 
where $\omega_1,\omega_2>0$, and $\theta>\delta$ are tuning parameters.  Here, \eqref{ETmechanismslow} is the slow ETM and \eqref{ETmechanismfast} is the fast ETM, with $d_z(t)$ and $d_{v,\varepsilon}(t)$ defined in \eqref{dzdv}.  The term $\mathrm{e}^{-2\theta t}$ ensures that the ETMs \eqref{ETmechanism} rule out Zeno behavior (see \cite{kang2021event}). 
{\color{black}The parameters $\omega_1$ and $\omega_2$ should be chosen such that LMIs (85) are feasible. In practice, they can first be selected sufficiently small and then gradually increased until the desired tradeoff between LMI feasibility and transmission frequency is achieved. Smaller values of $\omega_1$ and $\omega_2$ make the ETMs more sensitive to measurement changes and lead to more frequent transmissions, rendering the scheme closer to the continuous-time measurement case (cf. \cite{kang2021event}).}

Consider now the slow and fast time-triggered control sampling instants $0=t_0^u<t_1^u<\dots <t_k^u<\dots$ and $0=\varepsilon s_0^u<\varepsilon s_1^u< \dots < \varepsilon s_j^u < \dots$ satisfying $t_{k+1}^u-t_k^u \leq h_z$, $s_{j+1}^u-s_j^u \leq h_v$,  $\lim_{k\to \infty}t_k^u=\infty$, $\lim_{j\to \infty}s_j^u=\infty$.
Based on the observer \eqref{observerdyeq44SD}, we propose the following finite-dimensional sampled-data controller (see Fig. \ref{fig0}):   
\begin{equation}\label{SDobservercontrollera}
	\begin{split}
	&	{u}(t)=[u_1(t),u_2(t)]^{\mathrm{T}}=u_s(t)+u_f(t), \\
	&	u_s(t)=- K_1 \hat{z}^{N_0}(t_k^u), ~~t\in[t_k^u ,t^u_{k+1}), \\
	&	u_f(t)=-K_2 \hat{v}^{N_0}(\varepsilon s_j^u), ~~t\in[\varepsilon s_j^u , \varepsilon s_{j+1}^u),  
	\end{split}
\end{equation}
where $K_0=[K_{1}, K_{2}]$ with $K_{1},K_{2}\in\mathbb{R}^{2\times 2N_0}$ is the controller gain given by \eqref{KKss} (found from LMIs \eqref{LMItails} with $\varepsilon=0$ and \eqref{dyLMIa100ss}). 
Our objective is to derive LMI conditions for finding the maximum admissible values of $h_z$ and $h_v$ while preserving the exponential stability. 

For any fixed $\varepsilon>0$, the well-posedness of \eqref{eq1}, \eqref{observerdyeq44SD} under control law \eqref{SDobservercontrollera} can be established in a manner similar to Section \ref{WP} and \cite{katz2020boundary}. Let $\mathcal{J}_{\varepsilon}=\{\mathcal{J}_{j}^{\varepsilon}\}_{j=0}^{\infty} := \{t_k^u\}_{k=0}^{\infty}\cup \{t_k^y\}_{k=0}^{\infty}\cup \{\varepsilon s_j^u\}_{j=0}^{\infty} \cup \{\varepsilon s_j^y\}_{j=0}^{\infty}$ with $0=\mathcal{J}_{0}^{\varepsilon}<\mathcal{J}_{1}^{\varepsilon}<\dots$. We prove the well-posedness on $[\mathcal{J}_{j}^{\varepsilon},\mathcal{J}_{j+1}^{\varepsilon})$ step-by-step.  For $\varepsilon=0$, the fast samplings in measurement and control become the continuous-time ones. In this case, we let $\mathcal{J}=\{\mathcal{J}_{j}\}_{j=0}^{\infty} := \{t_k^u\}_{k=0}^{\infty}\cup \{t_k^y\}_{k=0}^{\infty}$ with $0=\mathcal{J}_{0}<\mathcal{J}_{1}<\dots$ and prove the well-posedness on $[\mathcal{J}_{j},\mathcal{J}_{j+1})$ step-by-step.  

Next, we give a lemma (see proof in Appendix \ref{App2}) to show that ETMs \eqref{ETmechanism} exclude Zeno behavior.
\begin{lemma}\label{Zone}
	The ETMs \eqref{ETmechanismslow} (slow) and \eqref{ETmechanismfast} (fast) exclude Zeno behavior, i.e.,  $\lim_{k\to\infty}t_k^y=\infty$ and $\lim_{j\to\infty}s_j^y=\infty$. 
\end{lemma} 

Based on Lemma \ref{Zone}, ETMs \eqref{ETmechanism} imply the following bounds:
\begin{equation}\label{ETmechanism11}
	\begin{split}
		& d_z^2(t) \leq  \omega_1 (|y_z(t)-\hat{y}_z(t)|^2  +  \mathrm{e}^{-2\theta t})\\
	 & = \omega_1(|C^0_z e^{N_0}_z(t)+C^1_z e^{N-N_0}_z(t)+\zeta_z(t)|^2+\mathrm{e}^{-2\theta t}), ~\forall t\geq 0,\\
	&	d_{v,\varepsilon}^2(t) \leq   \omega_2 (|y_v(t)-\hat{y}_v(t)|^2  + \mathrm{e}^{-2\theta t}) \\
	& = \omega_2(|C_v^0 e^{N_0}_{v}(t)+ C_v^1 e^{N-N_0}_{v}(t)+ \zeta_v(t)|^2+  \mathrm{e}^{-2\theta t}),~\forall t\geq 0. 
	\end{split}
\end{equation}

\begin{remark}
	For the sampled-data state-feedback control of SP ODEs, 
	the existing results assume that the slow and fast sampling intervals are synchronous (i.e., $\{s_j^u\}_{j=1}^{\infty}=\{t_k^u\}_{k=1}^{\infty}$ see \cite{djemai1999digital,fridman2006robust}), which is restrictive in practical applications. 
	In \cite{chen2020multi}, the asynchronous slow and fast sampling intervals were studied by introducing complicated Lyapunov functional. 
	In this paper, for the finite-dimensional part, we consider two-scale Wirtinger-based piecewise continuous in time Lyapunov functionals (see $V_z$ and $V_v$ in \eqref{SDsec5eq1ss}) that compensate the multi-rate controller samplings, leading to the simplest efficient stability conditions.
\end{remark}

Recalling the notations in \eqref{notationsN0}, \eqref{notation222}, we define    
 \begin{equation*}
	\begin{array}{rr}
		{\Gamma}(t)=\left[ \begin{array}{cc}
		{\Gamma}_z(t)\\
		{\Gamma}_v(t)
		\end{array} \right]=\left[ \begin{array}{cc}
		\hat{z}^{N_0}(t) - \hat{z}^{N_0}(t_k^u) \\
		\hat{v}^{N_0}(t) -\hat{v}^{N_0}( \varepsilon s_j^u)
		\end{array} \right],\\
		t\in [t_k^u,t_{k+1}^u)\cap [\varepsilon s_j^u, \varepsilon s_{j+1}^u).  
	\end{array}
\end{equation*}
Using  \eqref{observereq8aa}, \eqref{observerdyeq44SD}, \eqref{observereq8SD} and \eqref{SDobservercontrollera}, we obtain the following closed-loop system:  
\begin{align}\label{SDsec5eq1ss}
		&E_{N_0}^{\varepsilon}\dot{\hat{X}}_0(t)=({\bf A}_0-{\bf B}_0K_0) \hat{X}_0(t)+  {\bf B}_0K_0 \Gamma(t) \nonumber\\
		&~~~~~~+{\bf L}_0[{\bf C}_0e^{N_0}(t)+{\bf C}_1e^{N-N_0}(t) + \zeta(t)- {\bf d}_\varepsilon(t)], \nonumber\\
		&E_{N_0}^{\varepsilon}\dot{e}^{N_0}(t)=({\bf A}_0 -{\bf L}_0{\bf C}_0 )e^{N_0}(t) - {\bf L}_0{\bf C}_1e^{N-N_0}(t) \\
		&~~~~~~-{\bf L}_0\zeta(t) + {\bf L}_0{\bf d}_\varepsilon(t), \nonumber\\
		&E_1^{\varepsilon} \dot{X}_{n}(t) =A_{n}X_{n}(t) - b_{n} K_0[\hat{X}_0(t)-{\Gamma}(t) ],~n>N,  \nonumber
\end{align}
with $\zeta(t)$ subject to \eqref{zetazeta} and $e^{N-N_0}$ satisfying \eqref{eq225}. Moreover, ${\hat{X}}_1(t)$ satisfies  
\begin{equation}\label{eq224SD}
		E_{N-N_0}^{\varepsilon}\dot{\hat{X}}_1(t)={\bf A}_1 {\hat{X}}_1(t)-{\bf B}_1[\hat{X}_0(t)-{\Gamma}(t)].  
\end{equation}
Note that if $\hat{X}_0(t)$ is exponentially decaying, then $\hat{X}_0(t)-{\Gamma}(t)$ is also exponentially decaying. 
Therefore, by arguments similar to those in \eqref{eqX1}-\eqref{exstX1}, it follows that \eqref{eq224SD} is exponentially decaying (i.e., the solution satisfies \eqref{exstX1}). 
For the exponential stability of \eqref{eq225} and \eqref{SDsec5eq1ss}, we consider the Lyapunov functional 
\begin{align}
&	V_{\text{SD}}(t)= V(t) +V_{z}(t)+V_{v}(t), \nonumber \\
&	V_{z}(t)= h_z^2 \mathrm{e}^{2\delta h_z}\int_{t_k^u}^{t}\mathrm{e}^{-2\delta(t-s)}|\dot{\hat{z}}^{N_0}(s)|^2_{W_z}\mathrm{d}s  \nonumber \\
&~~   -\frac{\pi^2}{4}\int_{t_k^u}^{t}\mathrm{e}^{-2\delta(t-s)}|{\Gamma}_z(s)|^2_{W_z}\mathrm{d}s, ~t\in [t_k^u,t_{k+1}^u), \label{SDLyapunov} \\
&V_{v}(t)=(\varepsilon h_v)^2 \mathrm{e}^{2\delta \varepsilon h_v}\int_{\varepsilon s_j^u}^{t}\mathrm{e}^{-2\delta(t-s)}|\dot{\hat{v}}^{N_0}(s)|^2_{W_v}\mathrm{d}s \nonumber \\
&~~  -\frac{\pi^2}{4}\int_{\varepsilon s_j^u}^{t}\mathrm{e}^{-2\delta(t-s)}|{\Gamma}_v(s)|^2_{W_v}\mathrm{d}s , 	~t\in [\varepsilon s_j^u,\varepsilon s_{j+1}^u),\nonumber 
\end{align}
where $V(t)$ is defined in \eqref{SPLyapunovreduced} with $p>0$, $P_{0,\varepsilon}$, $P_{e,\varepsilon}$, $P_{1,\varepsilon}$ satisfying \eqref{PPP} and \eqref{SPineq1}, and $ 0<W_z,W_v \in \mathbb{R}^{N_0\times N_0}$. The term $V_{z}(t)$ compensates the slow input sampling, whereas  $V_{v}(t)$ compensates the fast input sampling.  
By Wirtinger's inequality (see Lemma 1 in \cite{selivanov2016observer}), one has $V_{z}(t)$ and $ V_{v}(t)\geq 0$. Moreover, $V_{z}(t_k^u)=0$, $k\in\mathbb{Z}_+$, and $V_{v}(\varepsilon s_j^{u})=0$, $j\in\mathbb{Z}_+$,  meaning that $V_{\text{SD}}(t)$ does not grow at the sampling instants. 

Differentiating $V_{\text{SD}}(t)$ along the solution to \eqref{SDsec5eq1ss}, we obtain  
\begin{align}\label{SD22observerabcd}
&		\dot{V}_{\text{SD}}(t)+2\delta V_{\text{SD}}(t) \nonumber\\
& =\hat{X}_0^{\mathrm{T}}(t) \Psi^{(11)}_{\varepsilon} \hat{X}_0(t)+2\hat{X}_0^{\mathrm{T}}(t)(P_{0}^{\varepsilon})^{\mathrm{T}} [{\bf B}_0K_0 \Gamma(t) \nonumber\\
& +{\bf L}_0{\bf C}_0e^{N_0}(t)+{\bf L}_0{\bf C}_1e^{N-N_0}(t) + {\bf L}_0\zeta(t)- {\bf L}_0{\bf d}_\varepsilon(t) ] \nonumber\\
& + (e^{N_0}(t))^{\mathrm{T}} \Psi^{(22)}_{\varepsilon}	e^{N_0}(t) 	 +(e^{N-N_0}(t))^{\mathrm{T}}\Psi^{(33)}_{\varepsilon}e^{N-N_0}(t) \nonumber\\
& + 2(e^{N_0}(t))^{\mathrm{T}} (P_{e}^{\varepsilon})^{\mathrm{T}}{\bf L}_0 [-{\bf C}_1e^{N-N_0}(t) -\zeta(t) +{\bf d}_\varepsilon(t)] \nonumber\\
&+	h_z^2\mathrm{e}^{2\delta h_z}|\dot{\hat{z}}^{N_0}(t)|^2_{W_z} -\frac{\pi^2}{4}|\Gamma_z(t)|^2_{W_z} \nonumber\\
& +  h_v^2\mathrm{e}^{2\delta \varepsilon h_v}|\varepsilon \dot{\hat{v}}^{N_0}(t)|^2_{W_v} -\frac{\pi^2}{4}|\Gamma_v(t)|^2_{W_v} \nonumber\\
&	+  \sum_{n=N+1}^{\infty} X_n^{\mathrm T}(t)[A_n+A_n^{\mathrm T} +2{\delta}E_1^{\varepsilon}]X_n(t)  \nonumber \\
&	- \sum_{n=N+1}^{\infty}2X_n^{\mathrm T}(t)b_n K_0[\hat{X}_0(t)-{\Gamma}(t) ],  
\end{align}
where $\Psi^{(11)}_{\varepsilon}$, $\Psi^{(22)}_{\varepsilon}$, and $\Psi^{(33)}_{\varepsilon}$ are defined in \eqref{psipsi}. 
By Young's inequality, we have for $0<R_1,R_2\in \mathbb{R}^{2\times 2}$, 
  \begin{align}\label{SDobserverdyeq2.15}
  &	-\sum_{n=N +1}^{\infty}2X_n^{\mathrm T}(t)b_n K_0[\hat{X}_0(t)- {\Gamma}(t) ] \nonumber \\
		\leq& \sum_{n=N+1}^{\infty} \lambda_n X_n^{\mathrm T}(t)(R_1^{-1}+R_2^{-1})X_n(t) \\
		&+ \kappa_{N}[\hat{X}_0^{\mathrm{T}}(t)K_0^{\mathrm{T}} R_1 K_0\hat{X}_0(t) + {\Gamma}^{\mathrm{T}}(t)K_0^{\mathrm{T}} R_2K_0{\Gamma}(t)], \nonumber
  \end{align}
where $\kappa_N$ is defined below \eqref{observerdyeq2.15}.

Let $\tilde{\eta}_0=\mathrm{col}\{\hat{X}_0, e^{N_0}, {\Gamma},  \zeta, {\bf d}_{\varepsilon}\}$ and $\tilde{\eta}=\mathrm{col}\{\tilde{\eta}_0, e^{N-N_0}\}$. 
Substituting \eqref{SDobserverdyeq2.15} into \eqref{SD22observerabcd}, using \eqref{zetazeta}, \eqref{ETmechanism11}, and S-procedure, we obtain for $a_1, a_2, a_3, a_4>0$,  
\begin{align}\label{SPineq2}
	& \dot{V}_{\text{SD}}(t)  +2\delta V_{\text{SD}}(t)  + a_1\left[ \|c_z\|^{2}_{N} \sum_{n=N+1}^{\infty} z^2_n(t) -|\zeta_z(t)|^{2}\right] \nonumber\\
&~~~~ + a_2\left[ \|c_v\|^{2}_{N} \sum_{n=N+1}^{\infty}v^2_n(t) -|\zeta_v(t)|^{2}\right]  \nonumber\\
&~~~~ + a_3 [ \omega_1 |C_z^0 e^{N_0}_z(t)+C_z^1 e^{N-N_0}_z(t)+\zeta_z(t)|^2  \\
&~~~~~ \quad\quad + \omega_1 \mathrm{e}^{-2\theta t} - d_z^{2}(t)  ] \nonumber\\
&~~~~ +a_4 [\omega_2 |C_v^0 e^{N_0}_{v}(t)+C_v^1 e^{N-N_0}_{v}(t)+\zeta_v(t)|^2 \nonumber\\
&~~~~~ \quad\quad + \omega_2 \mathrm{e}^{-2\theta t}  - d_v^2(t) ] \nonumber\\
& \leq \tilde{\eta}^{\mathrm{T}}(t) [\tilde{\Phi}_{\varepsilon}+\Xi^\mathrm{T} \Upsilon_\varepsilon W \Upsilon_\varepsilon \Xi ] \tilde{\eta}(t)   \nonumber\\
& ~~~~ +  \sum_{n=N+1}^{\infty} X_n^{\mathrm T}(t)\Theta^{\varepsilon}_{N+1}X_n(t) + (a_3\omega_1 +a_4\omega_2)\mathrm{e}^{-2\theta t} \nonumber\\
& \leq (a_3\omega_1 +a_4\omega_2) \mathrm{e}^{-2\theta t}, \nonumber
\end{align}
provided  
\begin{equation}\label{LMIsSDobserver}
\begin{split}
	&\tilde{\Phi}_{\varepsilon} = 	\left[\begin{array}{c|c}
		\Phi_{\varepsilon} &  ({\Psi}^{(16)}_{\varepsilon})^{\mathrm{T}} \\ \hline
		{\Psi}^{(16)}_{\varepsilon} & {\Phi}^{(66)}_{\varepsilon}
	\end{array}\right]  +\Xi^\mathrm{T} \Upsilon_\varepsilon W \Upsilon_\varepsilon \Xi   <0 ,  \\	
   & \tilde{\Theta}_{N+1}^{\varepsilon}= {\Theta}_{N+1}^{\varepsilon}+ \lambda_{N+1}  (R_1^{-1}+R_2^{-1}) <0 ,  
\end{split}	
\end{equation} 
where  
\begin{equation*}
	\begin{split}
&{\Theta}_{N+1}^{\varepsilon}= A_{N+1}+A_{N+1}^{\mathrm T} +2\delta E_{1}^{\varepsilon}  + \mathrm{diag}\{a_1\|c_z\|^{2}_{N}, a_2\|c_v\|^{2}_{N}\} ,\\
&\Phi_{\varepsilon} = \setlength{\arraycolsep}{1.5pt} {\left[ \begin{matrix}
	\Phi^{(11)}_{\varepsilon} & \Phi^{(12)}_{\varepsilon} & 
	(P_{0}^{\varepsilon})^{\mathrm{T}}{\bf B}_0K_0  & \Phi^{(14)}_{\varepsilon} & -(P_{0}^{\varepsilon})^{\mathrm{T}} {\bf L}_0  \\
	*& \Phi^{(22)}_{\varepsilon} & 0 & -(P_{e}^{\varepsilon})^{\mathrm{T}} {\bf L}_0 & (P_{e}^{\varepsilon})^{\mathrm{T}} {\bf L}_0  \\
	*&*& \Phi^{(33)} & 0 & 0  \\
	* & * & * & \Phi^{(44)} & 0 \\
	* & * & * & *& -\mathrm{diag}\{a_3, a_4\} 
\end{matrix} \right]} , \\
&\Phi^{(11)}_{\varepsilon}  = {\Psi}^{(11)}_{\varepsilon}+\kappa_{N}K_0^{\mathrm{T}} R_1 K_0  , ~\Phi^{(12)}_{\varepsilon}= (P_{0}^{\varepsilon})^{\mathrm{T}}{\bf L}_0 {\bf C}_0, \\
&\Phi^{(14)}_{\varepsilon}=(P_{0}^{\varepsilon})^{\mathrm{T}} {\bf L}_0+{\bf C}_0^{\mathrm{T}} \mathrm{diag}\{a_3\omega_1, a_4\omega_2  \}, \\
&\Phi^{(22)}_{\varepsilon}  = {\Psi}^{(22)}_{\varepsilon} + {\bf C}_0^{\mathrm{T}} 
\mathrm{diag}\{a_3\omega_1, a_4\omega_2  \}{\bf C}_0,  \\
&\Phi^{(33)}=-\frac{\pi^2}{4} W +\kappa_{N}K_0^{\mathrm{T}} R_2 K_0, 
W=\mathrm{diag}\{W_z,W_v\}, \\ 
&\Phi^{(44)}=\mathrm{diag}\{-a_1+a_3\omega_1, -a_2+ a_4\omega_2  \}, \\
&\Phi^{(66)}_{\varepsilon}=p[(P_{1}^{\varepsilon})^{\mathrm{T}}{\bf A}_1+{\bf A}_1^{\mathrm{T}}P_{1}^{\varepsilon}  +2\delta {E}_{N-N_0}^{\varepsilon} P_{1}^{\varepsilon} ] \\
&~~~~~~~~~ +{\bf C}_1^{\mathrm{T}} \mathrm{diag}\{a_3\omega_1, a_4\omega_2  \} {\bf C}_1, \\
&{\Phi}^{(16)}_{\varepsilon}= [((P_{0}^{\varepsilon})^{\mathrm{T}}{\bf L}_0 {\bf C}_1)^{\mathrm{T}},  -((P_e^{\varepsilon})^{\mathrm{T}}{\bf L}_0 {\bf C}_1)^{\mathrm{T}} , 0, 0, 0]  \\
&~~ +[0, {\bf C}_1^{\mathrm{T}}\mathrm{diag}\{a_3\omega_1, a_4\omega_2  \}{\bf C}_0 , 0, {\bf C}_1^{\mathrm{T}}\mathrm{diag}\{a_3\omega_1, a_4\omega_2  \}, 0 ], \\
&\Xi_{0}= [{\bf A}_0 -{\bf B}_0 K_0, ~  {\bf L}_0{\bf C}_0 , ~ {\bf B}_0K_0, ~  {\bf L}_0, ~ -{\bf L}_0  ],\\
&\Xi= [\Xi_{0}, {\bf L}_0{\bf C}_1 ], ~~\Upsilon_{\varepsilon}=\mathrm{diag}\{h_z\mathrm{e}^{\delta h_z}I,h_v\mathrm{e}^{\varepsilon \delta h_v }I\}.   
	\end{split} 
\end{equation*}
Let $P_{1}^{\varepsilon}$ satisfy \eqref{P1epsP1}. By choosing $p>0$ large enough and using Schur complement, we find that \eqref{LMIsSDobserver}  hold if  
\begin{equation}\label{eq3.14}
\begin{split}
&	\left[ \begin{array}{cc}
	\Phi_{\varepsilon}	& \Xi_{0}^\mathrm{T} \Upsilon_\varepsilon W  \\ 
	*& -W 
	\end{array}\right]<0,\\
& \left[\begin{array}{ccc}
		\tilde{\Theta}_{N+1}^{\varepsilon} & I & I\\
		*& -\frac{1}{\lambda_{N+1}} R_1 & 0 \\
		*&*& -\frac{1}{\lambda_{N+1}} R_2
		\end{array}\right]<0.   
\end{split}
\end{equation} 
Therefore, if LMIs \eqref{SPineq1}, \eqref{P1epsP1}, and \eqref{eq3.14} are feasible, we can obtain from \eqref{SPineq2}, the variation of constant formula, and the comparison principle that  
\begin{equation}\label{eq315}
	\begin{split}
		 V_{\text{SD}}(t)\leq&\, \mathrm{e}^{-2\delta t}V_{\text{SD}}(0)+\frac{ (a_3\omega_1 +a_4\omega_2) }{2(\theta-\delta)}\mathrm{e}^{-2\theta t}\\
		 \leq&\, M_{1}\mathrm{e}^{-2\delta t}, ~t\geq 0,  
	\end{split}
\end{equation}
for some  $M_{1}>0$. 
Combining \eqref{sanjiao}, \eqref{SDLyapunov}, \eqref{eq315} and the fact that the solution of \eqref{eq224SD} satisfies \eqref{exstX1}, we arrive at \eqref{expstable} for some $\varepsilon$-dependent $M(\varepsilon)>0$ and decay rate $\delta>0$. 

By arguments of Theorem \ref{thm4.1}, given $0=\varepsilon_0<\varepsilon_1$, inequalities \eqref{SPineq1}, \eqref{P1epsP1}, and \eqref{eq3.14} hold for all $\varepsilon\in (\varepsilon_0, \varepsilon_1]$, if 
\begin{align}\label{eq51SDSDobserver}
&{E}_{N_0}^{\varepsilon_1} P_{0}^{\varepsilon_1,1}>0, {E}_{N_0}^{\varepsilon_1} P_{e}^{\varepsilon_1,1}>0,  {E}_{N-N_0}^{\varepsilon_1} P_{1}^{\varepsilon_1,1}>0,  \nonumber \\		
&(P_{1}^{\varepsilon_i,i})^{\mathrm{T}}{\bf A}_1+{\bf A}_1^{\mathrm{T}}P_{1}^{\varepsilon_i,i}  +2\delta {E}_{N}^{\varepsilon_i} P_{1}^{\varepsilon_i,i} <0, ~~i=0,1,\nonumber  \\ 		
&\setlength{\arraycolsep}{1.5pt} {\small 	\left[ \begin{array}{c|c}
	\begin{matrix}
	\Phi^{(11)}_{\varepsilon_i,i} & \Phi^{(12)}_{\varepsilon_i,i} & \Phi^{(13)}_{\varepsilon_i,i} & (P_{0}^{\varepsilon_i,i})^{\mathrm{T}} {\bf L}_0 & - (P_{0}^{\varepsilon_i,i})^{\mathrm{T}} {\bf L}_0   \\
	*& \Phi^{(22)}_{\varepsilon_i,i} & 0 & -(P_{e}^{\varepsilon_i,i})^{\mathrm{T}} {\bf L}_0 & (P_{e}^{\varepsilon_i,i})^{\mathrm{T}} {\bf L}_0    \\
	*&*& \Phi^{(33)}  & 0 & 0    \\
	* & * & * & \Phi^{(44)} & 0  \\
	*&*&*&* & -\mathrm{diag}\{a_3, a_4\}
\end{matrix} & \Xi^{\mathrm{T}}_0\Upsilon_{\varepsilon_i}W \\ \hline
*& -W
	\end{array} \right]}<0, \nonumber  \\
& \left[\begin{array}{ccc}
		\tilde{\Theta}_{N+1}^{\varepsilon_i} & I & I\\
		*& -\frac{1}{\lambda_{N+1}} R_1 & 0 \\
		*&*& -\frac{1}{\lambda_{N+1}} R_2
		\end{array}\right]<0,   
 \end{align}
where  
\begin{equation*}
	\begin{split}
	 P_{\mu}^{\varepsilon_i,i} &\, = \left[\begin{array}{cc}
		P_{\mu,1}^{(i)} & \varepsilon_i P_{\mu,2}^{\mathrm{T}}\\
		P_{\mu,2} & P_{\mu,3}
		\end{array}\right], ~ \mu\in \{0,e,1\}, i=0,1, \\ 
		\Phi^{(11)}_{\varepsilon_i,i} &\,=	(P_{0}^{\varepsilon_i,i})^{\mathrm{T}}({\bf A}_0-{\bf B}_0K_0)+({\bf A}_0-{\bf B}_0K_0)^{\mathrm{T}}P_{0}^{\varepsilon_i,i}\\
	&\, +2\delta {E}_{N_0}^{\varepsilon_i} P_{0}^{\varepsilon_i,i} + \kappa_{N}{K}_0^{\mathrm{T}}R_1{K}_0 ,\\
	\Phi^{(12)}_{\varepsilon_i, i} &\, = (P_{0}^{\varepsilon_i,i})^{\mathrm{T}}{\bf L}_0 {\bf C}_0, ~ {\color{black}\Phi^{(13)}_{\varepsilon_i, i}=(P_{0}^{\varepsilon_i,i})^{\mathrm{T}}{\bf B}_0K_0}, \\ 
		\Phi^{(22)}_{\varepsilon_i,i} &\, = (P_{e}^{\varepsilon_i,i})^{\mathrm{T}} ({\bf A}_0-{\bf L}_0{\bf C}_0)+({\bf A}_0-{\bf L}_0{\bf C}_0)^{\mathrm{T}}P_{e}^{\varepsilon_i,i} \\
&\,  +2\delta {E}_{N_0}^{\varepsilon_i} P_{e}^{\varepsilon_i,i} +{\bf C}_0^{\mathrm{T}} 
\mathrm{diag}\{a_3\omega_1, a_4\omega_2  \}{\bf C}_0.  		
	\end{split}
\end{equation*}

Next, we show that LMIs \eqref{eq51SDSDobserver} are always feasible for small enough $\varepsilon_1, h_z,h_v,\omega_1,\omega_2>0$ and large enough $N$. It is sufficient to consider \eqref{eq3.14} with $\varepsilon=0$.  
We take $R_1=R_2=4I$, $W_z=\frac{1}{h_z}I$, $W_v=\frac{1}{h_v}I$, $a_3=\frac{1}{\omega_1}$, $a_4=\frac{1}{\omega_2}$, $a_1=\|c_z\|^{-2}_{N}$, and $a_2=\|c_v\|^{-2}_{N}$. Let $h_z, h_v,\omega_1,\omega_2 \to 0^+$ and $N\to \infty$, we find that \eqref{eq3.14} are feasible if  
\begin{equation}\label{eq3.18}
	\begin{split}
	&	A_{N+1}^{\delta}+(A_{N+1}^{\delta})^{\mathrm T}+  \frac{1}{2}\lambda_{N+1}I  + I <0,\\
	& \left[ \begin{matrix}
	{\Psi}^{(11)}_{0} & (P_{0}^{0})^{\mathrm{T}}{\bf L}_0 {\bf C}_0  \\
	*& {\Psi}^{(22)}_{0} + {\bf C}_0^{\mathrm{T}} {\bf C}_0 
\end{matrix} \right]  <0,  
	\end{split}
\end{equation}
where ${\Psi}^{(11)}_{0}$ and ${\Psi}^{(22)}_{0}$ are defined in \eqref{psipsi}. By following arguments similar to the proof of Proposition \ref{proposition2.1}, we can always find
\begin{equation*}
	\begin{split}
	P_{0}^{0}= \left[\begin{array}{cc}
		P_{0,1} & 0 \\
		P_{0,2} & P_{0, 3}
		\end{array}\right], ~~P_{0,1}>0,P_{0,3}>0,\\
	{P}_{e}^{0}= {\small\left[\begin{array}{cc}
		P_{e,1} & 0 \\
		P_{e,2} & P_{e,3}
		\end{array}\right]}, ~~P_{e,1}>0,P_{e,3}>0,	
	\end{split}
\end{equation*} 
such that \eqref{eq3.18} are feasible.   
Summarizing, we have  
\begin{theorem}\label{thm3.1}
	{\color{black}Consider system \eqref{eq1} with measurement \eqref{measurementSD} subject to the event-triggered mechanism \eqref{ETmechanism}}, observer \eqref{finobserver}, \eqref{observerdyeq44SD}, 
	sampled-data control law \eqref{SDobservercontrollera}, and initial value $z_0,v_0\in H^1(0,1)$. 
	Let $\delta>0$ be a desired decay rate, $N_0\in\mathbb{N}$ satisfy \eqref{NNN}, and $N\in\mathbb{N}$ satisfying $N\geq N_0$.
 Assume $c_z,c_v \in L^2(0,1)$ satisfy \eqref{cneq0}. 
	   Let the observer gain ${\bf L}_0$ be found from \eqref{observergain11} and given by \eqref{L0L0}.  Given $h_v, h_z>0$ and $ \varepsilon_1>0$, 
if there exist constants $a_1,a_2,a_3,a_4>0$, matrices $0<P_{0,1}^{(i)},P_{0,3}, P_{e,1}^{(i)}, P_{e,3} \in \mathbb{R}^{N_0 \times N_0}$, $0<P_{1,1}^{(i)}, P_{1,3}\in \mathbb{R}^{(N-N_0)\times (N-N_0)}$, $i=0,1$, 
 $P_{0,2}, P_{e,2} \in \mathbb{R}^{N_0 \times N_0}$,  $P_{1,2} \in \mathbb{R}^{(N-N_0) \times (N-N_0)}$, $0<W_z, W_v \in \mathbb{R}^{N_0\times N_0}$ and $0<R_1,R_2\in \mathbb{R}^{2\times 2}$ such that LMIs \eqref{eq51SDSDobserver} are feasible, 
 then solutions ${z}(x,t)$, ${v}(x,t)$ to \eqref{eq1} under the control law  \eqref{SDobservercontrollera} with $K_0$ given by \eqref{KKss} and the corresponding observer $\hat{z}(x,t)$, $\hat{v}(x,t)$ defined by  \eqref{finobserver} and \eqref{observerdyeq44SD} under the ETM \eqref{ETmechanism} satisfy \eqref{expstable} for all $\varepsilon\in (0 ,\varepsilon_1]$.  
	Moreover, {\color{black}LMIs \eqref{eq51SDSDobserver}} are always feasible for large enough $N$ and small enough $\varepsilon_1>0$, $h_z,h_v>0$.
\end{theorem}
\begin{remark}
	 For $\varepsilon=0$, our method allows sampling only in the slow variable. In this scenario, using the stability analysis in this section and following the arguments in Section \ref{sectionstability} with $\varepsilon=0$, we can   achieve exponential stabilization of descriptor PDE system \eqref{eq1slow}. 
	 Note that descriptor ODE system can be destabilized by arbitrary fast sampling in the fast variable of the state-feedback even if the system is stable under continuous-time state-feedback \cite{fridman2006robust}.
\end{remark}

\section{Numerical examples}\label{examples}
In this section, we consider system \eqref{eq1} under the following three sets of parameters:  
\begin{subequations}\label{numericalexamples}
\begin{align}
&\alpha=-8,~\beta=10, ~\rho=-3\pi, ~ \gamma=-\pi^2, \label{example1}\\
&\alpha=\pi,~\beta=\pi(\pi-1), ~\rho=-3\pi, ~ \gamma=-\pi,    \label{example2} \\
&  \alpha=5,~~\beta=6, ~~\rho=-10, ~~ \gamma=0.1. \label{example3}
\end{align}
\end{subequations}
{\color{black}For parameters \eqref{example1}, we have $\gamma=-\lambda_2$, which corresponds to a nonstandard SP system. This case cannot be handled in \cite{parada2024boundary} using a single boundary input, nor in \cite{morris2025boundary,parada2020feedback,wang2024finite} for $\varepsilon=0$.  The parameters \eqref{example2} also cannot be addressed  by \cite{parada2024boundary} via the backstepping approach since $\gamma<0$. Moreover, for \eqref{example2}, $\varepsilon=0$ leads to two unstable multiple eigenvalues, which is not treatable by the modal decomposition method with a single input (see \cite{wang2024finite}). 
The parameters \eqref{example3} are from \cite{parada2024boundary}. In this case, we show below that stabilization can be achieved using the single input $u_1$, but it may lead to high gain, whereas the introduction of an additional input $u_2$ can avoid high gains.}

First, we consider the continuous-time controller under continuous-time measurement \eqref{measurement} with  
\begin{equation}\label{czcv}
	c_z=\chi_{[0.1,0.5]}(x),~~ c_v=\chi_{[0.4,0.9]}(x),  
\end{equation}
where $\chi$ is the indicator function. 
Take $\delta=0.3$. The observer gain ${\bf L}_0$ is found from \eqref{observergain11}, \eqref{L0L0}:  
\begin{align}\label{observergainLL}
&{\bf L}_{(\mathrm{a})}=\left[\begin{matrix}
  5.9932  &    0.1305 &  -8.1235 &  -0.2142\\
  220.1382 &  5.9255 &   66.0505 &   1.2437
  \end{matrix} \right]^{\mathrm{T}},  \nonumber \\
&{\bf L}_{(\mathrm{b})}=\left[\begin{matrix}
   -58.013  & 175.070 & -3.441 &   3.813 \\
    85.942 &    -175.620 &   19.978 &  -0.36
         \end{matrix} \right]^{\mathrm{T}},\\        
&{\bf L}_{(\mathrm{c})}=\left[\begin{matrix}
    16.9874 &26.8380 & 4.0338 & 2.9491 \\
      45.2617 &  -1.3655 & 8.5991 &  2.2965
    \end{matrix} \right]^{\mathrm{T}}, \nonumber
\end{align}
where ${\bf L}_0={\bf L}_{(\mathrm{a})}$, ${\bf L}_{(\mathrm{b})}$, and ${\bf L}_{(\mathrm{c})}$ correspond to parameters \eqref{example1}, \eqref{example2}, and \eqref{example3}, respectively. 
Based on ${\bf L}_0$ in \eqref{observergainLL}, we verify the feasibility of LMIs \eqref{LMItails} with $\varepsilon=0$ and \eqref{dyLMIa100ss} for $N=10$ and obtain the control gains from \eqref{KKss}:  
\begin{equation}\label{controllergainKK}
	\begin{split}	
& K_{(\mathrm{a})}=\left[\begin{matrix}
35.9997  &  0 &   0.8171  &  0 \\
   -3.2566  & 0 &   24.6097 &   0
         \end{matrix} \right], \\	
&K_{(\mathrm{b})}=\left[\begin{matrix}
6.1062 &   -6.4444 &   0.5194  & -0.0183 \\
   39.9875 &  12.6107 &  10.6810  & -0.0049 
         \end{matrix} \right],\\      
&K_{(\mathrm{c})}= \left[\begin{matrix}
12.8574  &  0.5310   & 1.4451   &  -1.4675 \\
33.4467  &  16.7247  &  2.9653  &  0.3975
         \end{matrix} \right], 
\end{split}
\end{equation}
where $K_0=K_{(\mathrm{a})}$, $K_{(\mathrm{b})}$, and $K_{(\mathrm{c})}$ 
correspond to parameter sets \eqref{example1}, \eqref{example2}, and \eqref{example3}.   
{\color{black}The above examples also illustrate the role of the second actuator. For \eqref{example1}, $\gamma=-\lambda_2$, and the resulting nonstandard SP system is not covered by the single-input design in \cite{parada2024boundary} or by the descriptor-system results in \cite{morris2025boundary,parada2020feedback,wang2024finite}. For \eqref{example2}, the fast subsystem is unstable and the limiting descriptor system has multiple unstable eigenvalues; hence, the existing single-input designs are not applicable either.} 
{\color{black}For \eqref{example3}, the fast subsystem is stable and a single-input design is possible. Setting $u_2\equiv0$ and using the observer gain $L_0=L_{(c)}$, we obtain
\( K_{\mathrm{single}} =[7596.49,\;164.59,\;0,\;0].\)
This gain is considerably higher than the two-input gain $K_{(c)}$ in
\eqref{controllergainKK}. Therefore, the second actuator enlarges the class of systems covered by the proposed method and may also substantially reduce the controller gain.}

Substituting the gains \eqref{observergainLL} and \eqref{controllergainKK}, 
we verify the LMIs \eqref{SPineq1abc} for $N\in \{7,\dots,16\}$ to get the maximum $\varepsilon_1$ that preserves the feasibility. 
The results are given in Table \ref{table1}. 
\begin{table*}[!htb]
    \caption{Theorem \ref{thm4.1}: The maximum $\varepsilon_1$ that preserves the feasibility of LMIs.}\label{table1}
    \centering 
 \begin{tabular}{c|c|ccccccccccccc} \hline
 \centering Coefficients  & $N$ & 7 & 8 & 9 & 10 & 11 & 12& 13 & 14 & 15 & 16      \\ \cline{1-12} 
 \eqref{example1} &   $\varepsilon_1 $ & 0.17 & 0.34 & 0.50 & 0.75 & 0.90 & 1.12 & 1.26  & 1.39 & 1.64 & 1.78    \\ \hline 
 
 \eqref{example2} &   $\varepsilon_1 $ & 0.269 & 0.298 & 0.315  & 0.335 & 0.346 & 0.360  & 0.369 & 0.379 &0.391 & 0.398   \\ \hline 
 
\eqref{example3} & $\varepsilon_1 $  & 0.099 & 0.107 & 0.112 & 0.118 & 0.121 & 0.125 & 0.127 & 0.130 & 0.133 & 0.135    \\ \hline
\end{tabular}   
\end{table*}
{\color{black}Table~I shows that the maximum admissible SP parameter $\varepsilon_1$ generally increases with the observer dimension $N$. This is expected since increasing $N$ reduces the contribution of the residual modes and hence the conservatism of the LMI conditions.}

\begin{figure}[!htb]
\centering
\includegraphics[width=9.5cm]{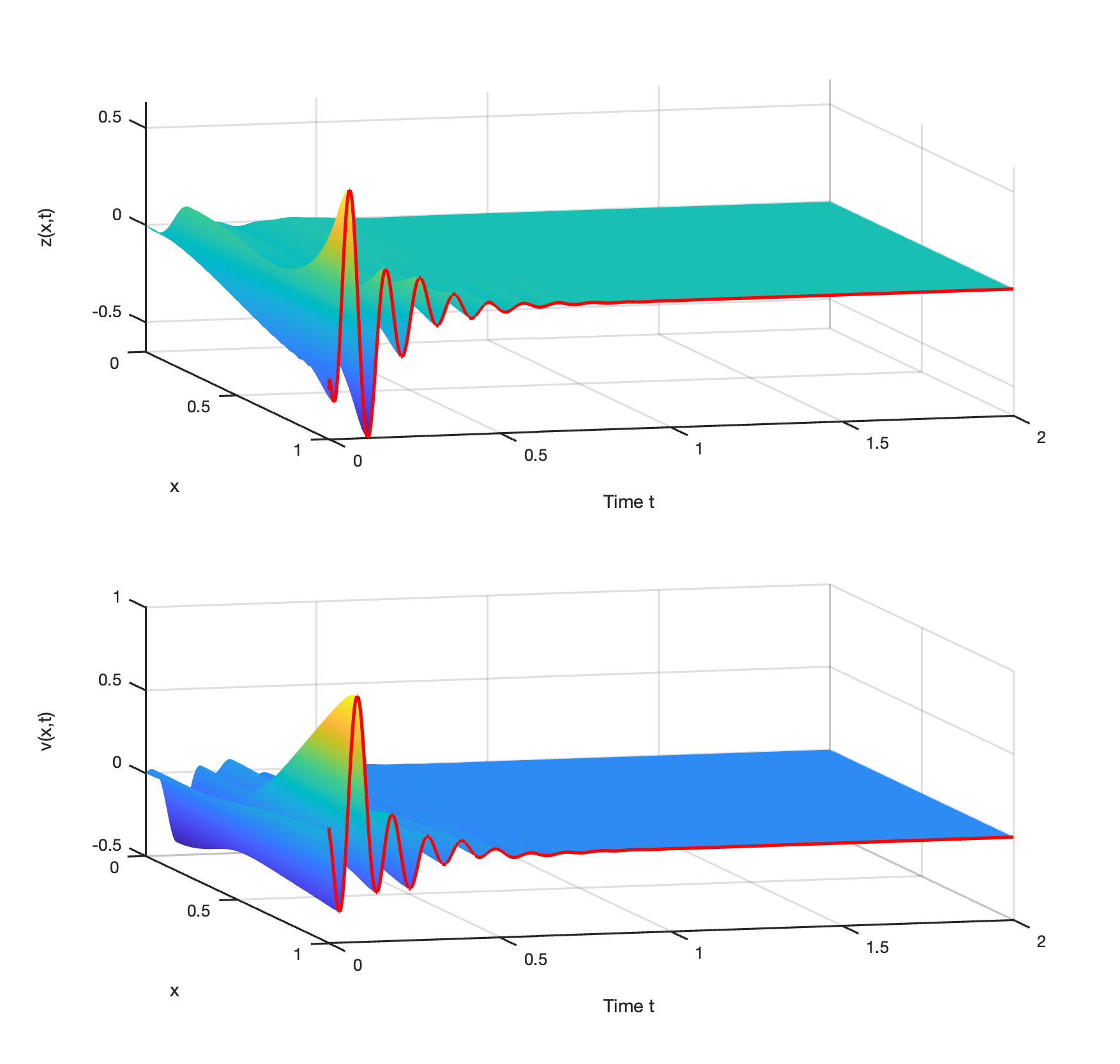} 
\caption{Evolutions of $z(x,t)$ (above) and $v(x,t)$ (below) for parameters \eqref{example1} with $N=8$ and $\varepsilon=0.34$.}\label{fig1}
\end{figure}

For the simulations, we choose the initial conditions $z_0(x)=\frac{2}{3}x^3-x^2$, $v_0(x)=-\frac{1}{3}x^3+\frac{1}{2}x^2$. 
We consider the parameter set \eqref{example1} with gains $\mathbf{L}_0=\mathbf{L}_{(a)}$, $K_0=K_{(a)}$, $N_0=2$, $N=8$, $\varepsilon=0.34$ (selected from Table \ref{table1}).  
The simulation was carried out using the approximation $z(x,t)\approx \sum_{n=1}^{100}z_n(t)\phi_n(x)$ and $v(x,t)\approx \sum_{n=1}^{100}v_n(t)\phi_n(x)$, together with the time discretization $X_n((i+1)\Delta) \approx \mathrm{e}^{E_{1}^{\frac{1}{\varepsilon}}A_n\Delta }X_n(i\Delta)+\Delta \mathrm{e}^{E_{1}^{\frac{1}{\varepsilon}}A_n\Delta } E_{1}^{\frac{1}{\varepsilon}} B_n{u}(i\Delta t)$ using a time step $\Delta= 0.001$ for each mode. The simulation results are presented in Fig. \ref{fig1} and confirm the theoretical results. 

For the sampled-data implementation, consider the sampled-data measurement \eqref{measurementSD} with $c_z$ and $c_v$ as in \eqref{czcv}. 
Let the update of the measurement sampling instants be determined by the event-triggered mechanism \eqref{ETmechanism} with $\omega_1=\omega_2=0.001$, 
and $\theta=1$. 
Consider the parameters \eqref{example1} with $\varepsilon=0.1$ and gains ${\bf L}_0 ={\bf L}_{(\mathrm{a})}$, $K_0=K_{(\mathrm{a})}$, 
parameters \eqref{example2} with $\varepsilon=0.1$ and gains ${\bf L}_0 ={\bf L}_{(\mathrm{b})}$, $K_0=K_{(\mathrm{b})}$, 
and parameters \eqref{example3} with $\varepsilon=0.05$ and gains ${\bf L}_0 ={\bf L}_{(\mathrm{c})}$ and $K_0=K_{(\mathrm{c})}$, respectively. 
For all cases, we fix $h_z=0.01$ for finding $h_v^{\max}$ (the maximal admissible $h_v$) and fix $h_v=0.01$ for finding $h_z^{\max}$ (the maximal admissible $h_z$), respectively. The results are presented in Table \ref{table2}. 
{\color{black}Table~\ref{table2} shows that the maximum admissible sampling intervals $h_z^{\max}$ and $h_v^{\max}$ increase as $N$ increases.  Moreover, the different admissible values of $h_z^{\max}$ and $h_v^{\max}$ illustrate the advantage of the proposed multi-rate implementation, which allows the slow and fast variables to be sampled at different rates.}
\begin{table*}[!htb]
    \caption{{\color{black}Theorem \ref{thm3.1}}: $w_1=w_2=0.001$ ($\varepsilon=0.1$ for parameters \eqref{example1} and \eqref{example2} and $\varepsilon=0.05$ for parameters \eqref{example3}).}\label{table2}
    \centering 
\begin{tabular}{c|c|ccccccccccccc} \hline
 \centering System parameters  & $N$ & 7 & 8 & 9 & 10 & 11 & 12& 13 & 14 & 15 & 16      \\ \cline{1-12} 
 \eqref{example1} with $h_v=0.01$ & $h_z^{\max}$ & 0.007 & 0.023 & 0.027 & 0.030 & 0.032 & 0.033 & 0.034  & 0.035 & 0.036 & 0.037    \\ \hline 
  \eqref{example1} with $h_z=0.01$ & $h_v^{\max}$ & 0.011 & 0.042 & 0.048 & 0.053 & 0.055 & 0.056 & 0.057  & 0.058 & 0.059 & 0.060    \\ \hline 
  
   \eqref{example2} with  $h_v=0.01$ &   $h_z^{\max}$ & 0.020   &  0.025 &  0.029  & 0.033 & 0.035 & 0.038  & 0.040 & 0.042 &0.045 & 0.047   \\ \hline 
 
 \eqref{example2} with $h_z=0.01$ &   $h_v^{\max}$ & 0.104  &  0.119  &    0.124 & 0.130 & 0.132 & 0.134  & 0.136 & 0.137 &0.138 & 0.139   \\ \hline 

\eqref{example3} with $h_z=0.01$ & $h_v^{\max}$  & 0.003 & 0.045 & 0.068 & 0.093 & 0.106 & 0.121 & 0.129 & 0.138 & 0.148 & 0.153    \\ \hline
\eqref{example3} with $h_v=0.01$ & $h_z^{\max}$  & 0.009 & 0.011 & 0.012 & 0.013 & 0.014 & 0.014 & 0.015 & 0.015 & 0.016 & 0.016    \\ \hline
\end{tabular}    
\end{table*}

\begin{figure}[hbt!]
\centering
\includegraphics[width=9.5cm]{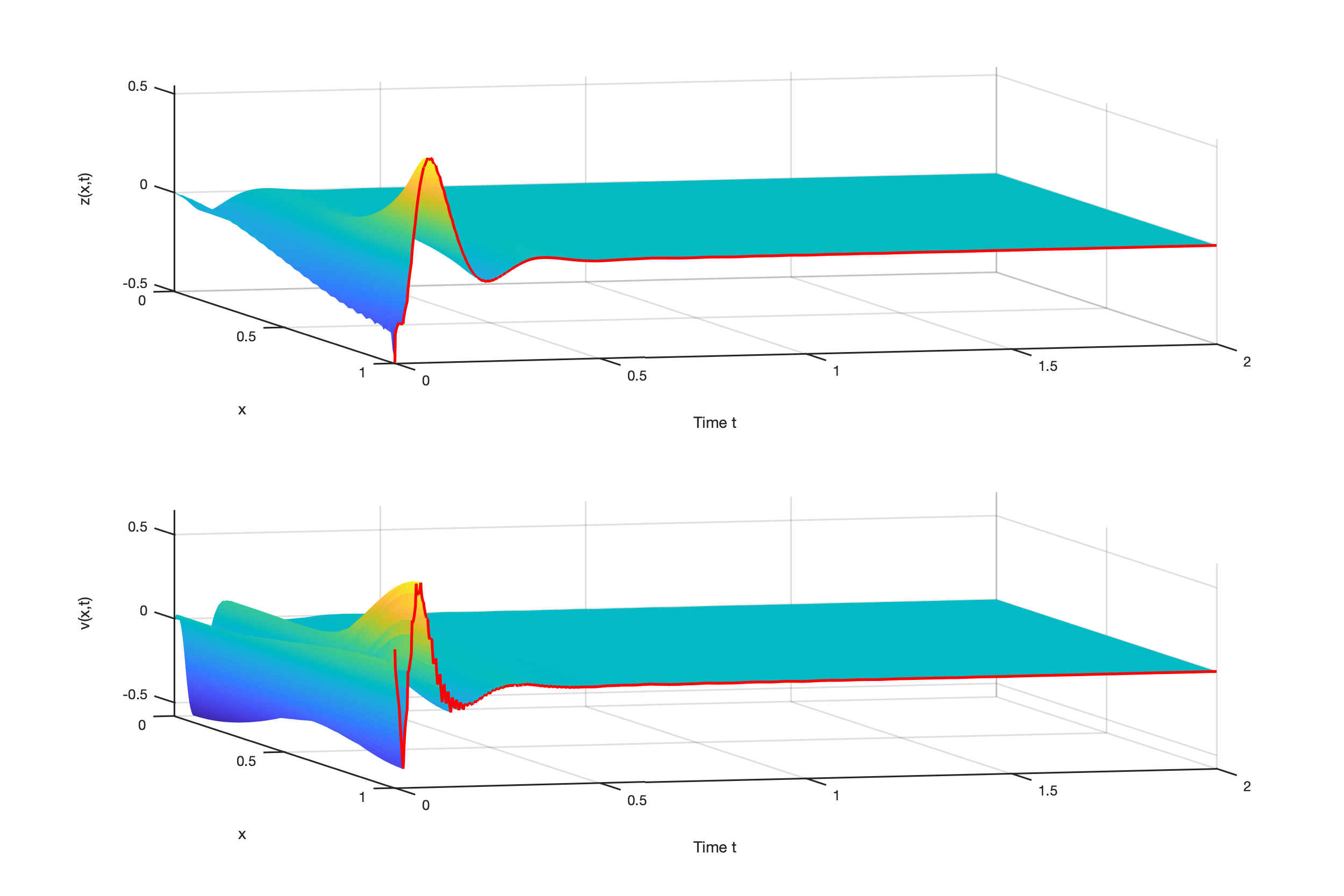} 
\caption{Evolutions of $z(x,t)$ (above) and $v(x,t)$ (below) for parameters \eqref{example2} with $N=10$, $\varepsilon=0.1$,  $h_z=0.01$, $h_v=0.13$. }\label{fig2}
\end{figure}

\begin{figure}[hbt!]
\centering
\includegraphics[width=9cm]{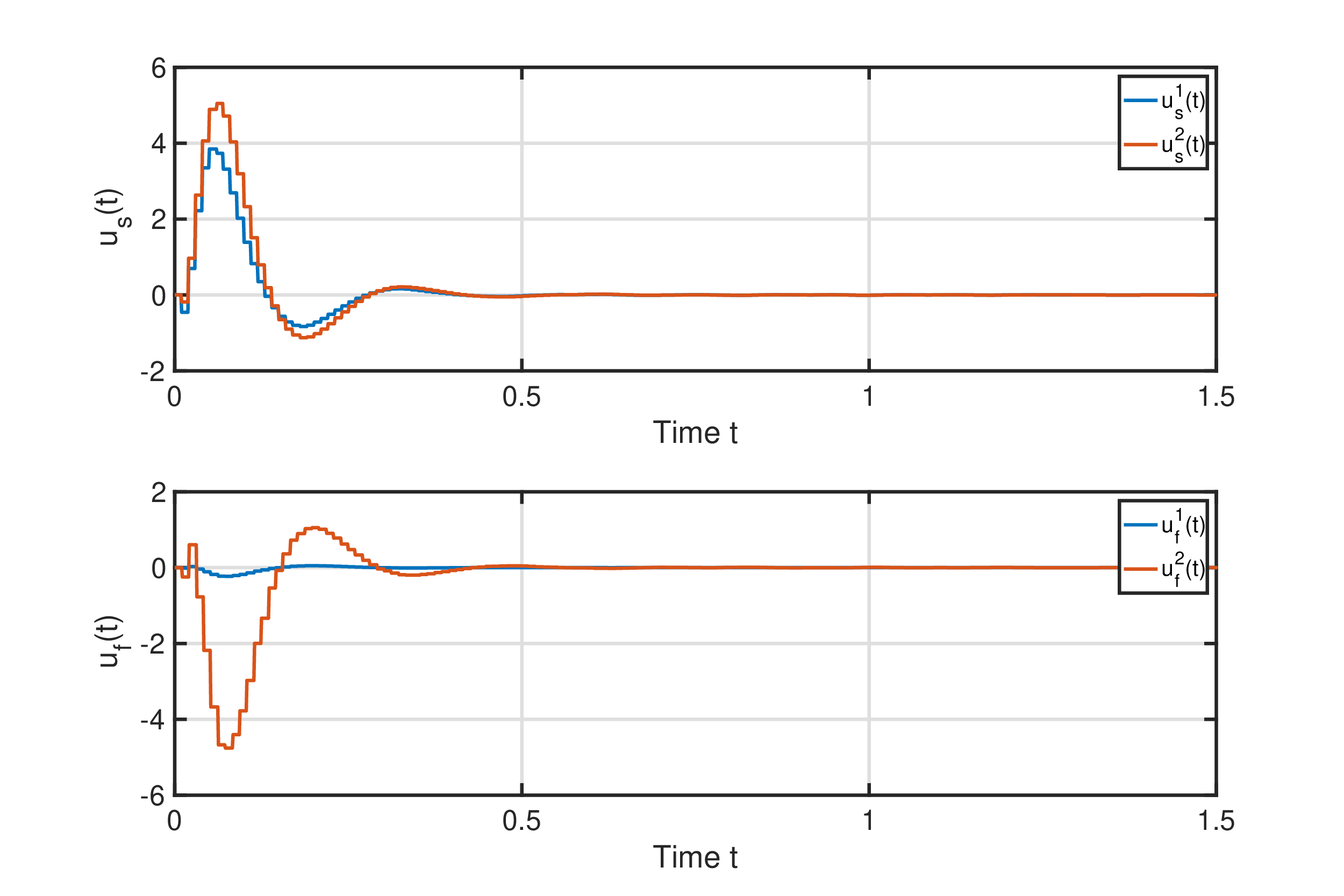} 
\caption{Evolutions of $u_s(t)$ (above) and $u_f(t)$ (below) for parameters \eqref{example2} with $N=10$, $\varepsilon=0.1$,  $h_z=0.01$, $h_v=0.13$. }\label{fig3}
\end{figure}

\begin{figure}[hbt!]
\centering
\includegraphics[width=9cm]{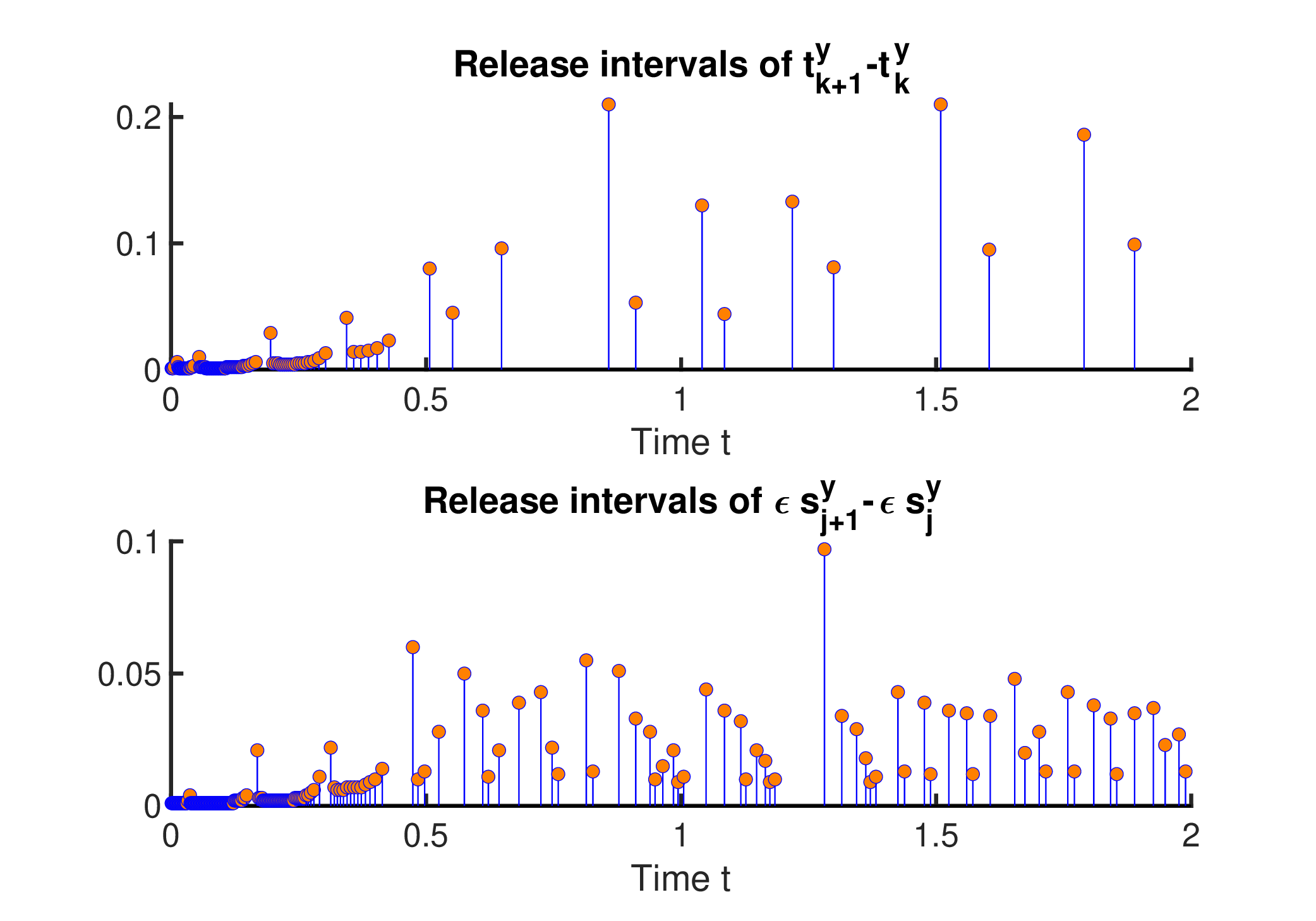} 
\caption{Release instants and release intervals for parameters \eqref{example2} with $\varepsilon=0.1$, $h_z=0.01$, $h_v=0.13$: $t_{k+1}^y-t_{k}^y$ (above) and $\varepsilon s_{j+1}^y- \varepsilon s_{j}^y$ (below).}\label{fig4}
\end{figure}

For simulations, we consider parameters \eqref{example2} with $\varepsilon=0.1$, $N=10$, $h_z=0.01$, $h_v=0.13$ (found from Table \ref{table2}). 
The simulation results in Figs. \ref{fig2}, \ref{fig3}, and \ref{fig4} confirm the theoretical results.  

\section{Conclusions}\label{secV} 
In this paper, we considered the finite-dimensional boundary observer-based control of a class of linear nonstandard SP reaction-diffusion systems. Both the continuous-time controller and its multi-rate sampled-data implementation were studied. We provided constructive LMI conditions for finding the upper bound of the perturbation parameter, observer dimension, and sampling intervals. Extensions to the control of various SP PDEs may be topics for future research.

\appendix

\subsection{Proof of Proposition \ref{propoWP}}\label{proofpropoWP} 
First, we represent $\mathcal{M}_{11} = \mathcal{T} + \tilde{\psi}f$, 
where $\mathcal{T} := -\mathcal{A}-\gamma \mathcal{I}$ and $f:=\tilde{\mathcal{K}}_{22}P_N=\sum_{n=1}^{N_0} (\tilde{\mathcal{K}}_{22})_n \langle \cdot, \phi_n \rangle$, where $(\tilde{\mathcal{K}}_{22})_n$ means the $n$-th entry of $\tilde{\mathcal{K}}_{22}$. 
Since $\gamma = -\lambda_{n_0}$ for some $n_0 \in \{1,\dots,N_0\}$, $\mathrm{ker} \mathcal{T} = \operatorname{span}\{\phi_{n_0}\}$ and $\mathcal{T}$ is invertible on $\phi_{n_0}^\perp:=\mathrm{span}\{\phi_n,n\neq n_0\}$. 

Assume $\mathcal{M}_{11} w = 0$. Decompose $w = \alpha\phi_{n_0} + w_\perp$ with $w_\perp \in \phi_{n_0}^\perp$. Then
\begin{equation}\label{appendix1}
    0 = \mathcal{M}_{11} w = \mathcal{T} w_\perp + \tilde{\psi}[\alpha(\tilde{\mathcal{K}}_{22})_{n_0} + f(w_\perp)]. 
\end{equation}
Taking inner product with $\phi_{n_0}$ and using $\langle \mathcal{T} w_\perp, \phi_{n_0}\rangle= \langle \mathcal{T} \phi_{n_0}, w_\perp \rangle=0$, we have $ \langle \tilde{\psi}, \phi_{n_0} \rangle [(\tilde{\mathcal{K}}_{22})_{n_0}\alpha + f(w_\perp)] = 0$, which together with $\langle \tilde{\psi}, \phi_{n_0} \rangle\neq 0$ implies
\begin{equation}\label{appendix2}
    (\tilde{\mathcal{K}}_{22})_{n_0}\alpha + f(w_\perp)=0.
\end{equation}
Substituting \eqref{appendix2} into \eqref{appendix1}, we get $\mathcal{T} w_\perp=0$, hence $w_\perp = 0$ and $f(w_\perp)=0$. 
Inequality \eqref{Lyapunoveq1} implies that $A_{\gamma,0} - B_0 K_{22}$ is Hurwitz, and thus $(K_{22})_{n_0} \neq 0$, otherwise, the $n_0$-th column of $A_{\gamma,0} - B_0 K_{22}$ would be the zero vector, which contradicts the Hurwitz property. 
This together with  \eqref{appendix2} implies $\alpha=0$. Thus $w=0$, proving that $\mathcal{M}_{11}$ is injective.

For arbitrary $h \in L^2(0,1)$, write $h = \beta\phi_{n_0} + h_\perp$ with $h_\perp \in  \phi_{n_0}^{\perp}$. Seek $w = \alpha\phi_{n_0} + w_\perp$ solving $\mathcal{M}_{11} w = h$:
\begin{equation}\label{appendixa2}
    \mathcal{T} w_\perp + \tilde{\psi}\big[\alpha(\tilde{\mathcal{K}}_{22})_{n_0} + f(w_\perp)\big] = \beta\phi_{n_0} + h_\perp. 
\end{equation}
Taking inner product with $\phi_{n_0}$, we obtain
$\langle \tilde{\psi}, \phi_{n_0} \rangle \big[(\tilde{\mathcal{K}}_{22})_{n_0}\alpha + f(w_\perp)\big] = \beta$, hence
\begin{equation}\label{appendixa3}
    (\tilde{\mathcal{K}}_{22})_{n_0}\alpha + f(w_\perp) = {\beta}{\langle \tilde{\psi}, \phi_{n_0} \rangle}^{-1}.
\end{equation}
Let $P_{\perp}$ be the orthogonal projection onto $\phi_{n_0}^\perp$. Since $\mathcal{T} w_\perp \in \phi_{n_0}^\perp$ and $h_\perp \in \phi_{n_0}^\perp$, we obtain from \eqref{appendixa2} and \eqref{appendixa3} that $\mathcal{T} w_\perp + \beta \langle \tilde{\psi}, \phi_{n_0} \rangle^{-1} P_{\perp}\tilde{\psi} = h_\perp$. 
Since $\mathcal{T}$ is invertible on $\phi_{n_0}^\perp$, we have $w_\perp = \mathcal{T}^{-1}(h_\perp - \beta \langle \tilde{\psi}, \phi_{n_0} \rangle^{-1} P_{\perp}\tilde{\psi})$. 
With $w_\perp$ determined, we compute $f(w_\perp)$ and then obtain $\alpha$ from \eqref{appendixa3} that $\alpha = (\tilde{\mathcal{K}}_{22})_{n_0}^{-1}(\beta \langle \tilde{\psi}, \phi_{n_0} \rangle^{-1} - f(w_\perp))$.  
Thus a unique solution $w = \alpha\phi_{n_0} + w_\perp$ exists for every $h$, proving that $\mathcal{M}_{11}$ is surjective. Therefore, $\mathcal{M}_{11}$ is bijective. Since $\mathcal{M}_{11}$ is closed, $\mathcal{M}_{11}^{-1}: L^2(0,1) \to L^2(0,1)$ is bounded by the Closed Graph Theorem. 

Second, write $\mathcal{M}_{22} = \operatorname{diag}\{M_1,M_2\}$ with $M_1 = A_{\gamma,0}-\mathbf{L}_0^{(22)}C_{v,0}-\psi_{c_v}^{N}\mathbf{L}_0^{(22)}\tilde{K}_{22}$ and $M_2 = \operatorname{diag}\{-\lambda_n-\gamma\}_{n=N_0+1}^N$. The matrix $A_{\gamma,0}-\mathbf{L}_0^{(22)}C_{v,0}$ is Hurwitz (see below \eqref{L0L0}), so $M_1$ is invertible for sufficiently large $k$. From \eqref{fasteigenvalues}, $M_2$ is diagonal with positive entries. Thus $\mathcal{M}_{22}$ is invertible. 

Finally, for a block operator with invertible diagonal blocks, $\mathcal{M}_0$ is invertible if and only if its Schur complement $S := \mathcal{M}_{22} - \mathcal{M}_{21}\mathcal{M}_{11}^{-1}\mathcal{M}_{12}$ is invertible. Since $\mathcal{M}_{12}x = -\tilde{\psi}\tilde{\mathcal{K}}_{22}x$ has rank one, $S$ is a rank-one perturbation of the invertible matrix $\mathcal{M}_{22}$. Choosing $\psi$ such that $1 + \tilde{\mathcal{K}}_{22} \mathcal{M}_{22}^{-1}(\mathcal{M}_{21}\mathcal{M}_{11}^{-1}\tilde{\psi}) \neq 0$ guarantees $S$ is invertible (Sherman-Morrison \cite{sherman1950adjustment}). With $\mathcal{M}_{11}$ and $S$ invertible, the block matrix inversion formula gives an explicit bounded inverse for $\mathcal{M}_0$. Therefore, $\mathcal{M}_0$ is boundedly invertible.

\subsection{Proof of Proposition \ref{propos2.1}}
\label{proofProposition} 
To prove the result, it is sufficient to show the feasibility of $\tilde{\Psi}_0<0$ (defined in \eqref{LMIcont}) and $\Theta_{N+1}^{0}<0$ (defined in \eqref{LMItails}) for sufficiently large $N$. 
By selecting $R=\frac{1}{2}I_2 $, $a_1 = \|c_z\|^{-2}_{N}$, and
 $a_2=\|c_v\|^{-2}_{N}$, we find that $\Theta_{N+1}^{0}<0$ always holds for sufficiently large $N$.  

We next prove $\tilde{\Psi}_0<0$. From \eqref{Lyapunoveq1}, for a given $\chi_1>0$ that is sufficiently small and independent of $N$, we can find an $N$-independent $P_{0}^{0}= {\small \left[\begin{array}{cc}
		P_{0,1} & 0 \\
		P_{0,2} & P_{0, 3}
		\end{array}\right]}$, $P_{0,1}>0,P_{0,3}>0$ such that  
\begin{equation}\label{psi11}
		\Psi^{(11)}_{0}= (P_{0}^{0})^{\mathrm{T}} A_{K_0}  +A_{K_0}^{\mathrm{T}}P_{0}^{0} < - \chi_1 I.   
\end{equation} 
For the fixed $\chi_1>0$, we take $\chi_2>0$ satisfying  
\begin{equation}\label{eq4.30}
		-\chi_2 I+\frac{1}{\chi_1}({P}_{0}^{0})^{\mathrm{T}}{\bf L}_0 {\bf C}_0 {\bf C}_0^{\mathrm{T}}{\bf L}_0^{\mathrm{T}} {P}_{0}^{0}<0.  
\end{equation}  
Since $(E_{N_0}, ({\bf A}_0^{\delta})^{\mathrm{T}},{\bf C}^{\mathrm{T}}_0)$ is stabilizable and the pair $(A_{\gamma,0}+\delta I, [0_{1\times N_0}, C_{v,0}])$ is controllable (by \eqref{cneq0} and Hautus lemma), we can choose ${\bf L}_0$ with
\begin{equation*}
        A_{{\bf L}_0}=\left[\begin{array}{ccc}
		A_{11}& A_{12} \\
        {A}_{21} & {A}_{22} 	\end{array}\right]={\bf A}_0^{\delta} -{\bf L}_0{\bf C}_0,	  
\end{equation*}
such that both $A_{22}$ and $A_s:=A_{11}-A_{12}A_{22}^{-1}A_{21}$ are Hurwitz (see the proof of Lemma 3.2 in \cite{khalil1989feedback}).
We claim that for any $\chi_2>0$ satisfying \eqref{eq4.30}, there exists an $N$-independent ${P}_{e}^{0}= {\small \left[\begin{array}{cc}
		P_{e,1} & 0 \\
		P_{e,2} & P_{e,3}
		\end{array}\right]}$, $0<P_{e,1}, P_{e,3}\in \mathbb{R}^{N_0\times N_0}$ such that  
\begin{equation}\label{eq4.31}
		\Psi^{(22)}_{0}=({P}_{e}^{0})^{\mathrm{T}}A_{{\bf L}_0}+A_{{\bf L}_0}^{\mathrm{T}}{P}_{e}^{0} <-\chi_2 I.  
\end{equation}
 We present $\Psi^{(22)}_{0}$ as   
	\begin{equation*}
	\begin{split}
	&\Psi^{(22)}_{0}={\small\left[\begin{array}{ccc}
		F_{11}& F_{12} \\
        * & F_{22} 	\end{array}\right]},~~ F_{22} = P_{e,3} A_{22} + A_{22}^{\mathrm{T}} P_{e,3}, \\
	&F_{11}=A_{11}^{\mathrm{T}}P_{e,1} + P_{e,1} A_{11}+ P_{e,2}^{\mathrm{T}}A_{21}+A_{21}^{\mathrm{T}}P_{e,2},\\
	&F_{12}= P_{e,1} A_{12} + P_{e,2}^{\mathrm{T}} A_{22} + A_{21}^{\mathrm{T}}P_{e,3}.  
	\end{split}
\end{equation*}
Since $A_{22}$ is Hurwitz, we can first choose $P_{e,3}>0$ such that $F_{22}=-\chi_2 I$. 
Let ${P}_{e,2}=-A_{22}^{-\mathrm{T}}(A_{12}^{\mathrm{T}}P_{e,1} +P_{e,3}A_{21})$, where $P_{e,1}$ is to be determined later. We have $F_{12}=0$. 
Then substituting the above ${P}_{e,2}$ into $F_{11}$, we arrive at  
\begin{equation*}
	\begin{split}
		&F_{11}=A_{11}^{\mathrm{T}} {P}_{e,1} + {P}_{e,1} A_{11}- (A_{12}^{\mathrm{T}} {P}_{e,1} + {P}_{e,3}A_{21})^{\mathrm{T}}A_{22}^{-1}A_{21}\\
		&\quad  -A_{21}^{\mathrm{T}}A_{22}^{-\mathrm{T}}(A_{12}^{\mathrm{T}}{P}_{e,1} + {P}_{e,3}A_{21})\\
		&=A_s^{\mathrm{T}} {P}_{e,1}+ {P}_{e,1}A_s-A_{21}^{\mathrm{T}}A_{22}^{-\mathrm{T}}{P}_{e,3}A_{21} -  A_{21}^{\mathrm{T}}{P}_{e,3}A_{22}^{-1}A_{21}.  
	\end{split}
\end{equation*}
Choose $\tilde{\chi}_2$ satisfying $\tilde{\chi}_2>\chi_2$ and $\tilde{\chi}_2 I > A_{21}^{\mathrm{T}}P_{e,3}A_{22}^{-1}A_{21}+(A_{21}^{\mathrm{T}}P_{e,3}A_{22}^{-1}A_{21})^{\mathrm{T}} $. Since $A_s$ is Hurwitz, there exists $P_{e,1}>0$ such that   
\begin{equation*}
	\begin{split}
&	A_s^{\mathrm{T}}{P}_{e,1}+{P}_{e,1} A_s\\
&	=-\tilde{\chi}_1 I+A_{21}^{\mathrm{T}}{P}_{e,3}A_{22}^{-1}A_{21}+(A_{21}^{\mathrm{T}}{P}_{e,3}A_{22}^{-1}A_{21})^{\mathrm{T}},   
	\end{split}
\end{equation*}	
which implies $F_{11}=-\tilde{\chi}_1 I<-\chi_2 I$. This, together with $F_{12}=0$ and $F_{22}=-\chi_2 I$, implies \eqref{eq4.31}. 
Combining \eqref{psi11}, \eqref{eq4.30}, and \eqref{eq4.31}, we conclude that   
\begin{equation}\label{eq63aa}
	\left[ \begin{matrix}
	{\Psi}^{(11)}_{0} & (P_{0}^{0})^{\mathrm{T}}{\bf L}_0 {\bf C}_0 \\
	*& \Psi^{(22)}_{0} 
\end{matrix} \right]<0.  
\end{equation}
Note that both $P_{0}^0$ and $P_e^0$ are independent of $N$. By using \eqref{eq63aa} and substituting $R=\frac{1}{2}I_2 $, $a_1 = \|c_z\|^{-2}_{N}$, $a_2=\|c_v\|^{-2}_{N}$, and the above $P_{0}^0$ and $P_e^0$ into $\tilde{\Psi}_0$ defined in \eqref{LMIcont}, we find that $\tilde{\Psi}_0<0$ for sufficiently large $N$. 
By continuity, it follows that the LMIs \eqref{SPineq1abc} are feasible for sufficiently small $\varepsilon_1>0$ and sufficiently large $N\in \mathbb{N}$. 

\subsection{Proof of Lemma \ref{Zone}}\label{App2} 
Consider first the slow sampling error $d_z(t)$ on an arbitrary finite interval $[0,T]$ with $T>0$. 
Since $d_z(t)$ is continuous on $[t_k^y, t_{k+1}^y) \subset [0,T]$, 
it follows that for any $\sigma>0$, there exists $\tau^* > 0$ such that for all $t, s \in [t_k^y, t^y_{k+1})  \subset [0,T] $,  
\begin{equation*}
	|t-s|<\tau^* \quad \Longrightarrow \quad |d_z(t)-d_z(s)|<\sigma.  
\end{equation*} 
Thus, the following reasoning by contraposition holds:  
\begin{equation*}
	\begin{split}
	&	\forall \sigma>0, \exists \tau^*>0, ~\forall t,s\in [t_k^y,t_{k+1}^y)\subset [0, T],\\
	&	|d_z(t)-d_z(s)|\ge  \sigma \Longrightarrow |t-s|\ge \tau^*.  
	\end{split}
\end{equation*}
Since $d_z(t_k^y)=0$, we have  
\begin{equation*}
	\begin{split}
	&	|d_z((t_{k+1}^{y})^{-})|=|d_z((t_{k+1}^{y})^{-})-d_{z}(t^{y}_k)|\\
	&	\geq \omega_1 (|y_z(t^{y}_{k+1}) -  \hat{y}_z(t^{y}_{k+1}) |^2 +  \mathrm{e}^{-2\theta t_{k+1}}) >\omega_1 \mathrm{e}^{-2\theta T}.  
	\end{split}
\end{equation*}
Thus, taking $\sigma(T)=\omega_1 \mathrm{e}^{-2\theta T}$ in the continuity argument above guarantees that there exists $\tau^*(T)>0$ such that $|t^y_{k+1}-t^y_k|\ge \tau^*(T)$. 
To rule out Zeno behavior on $[0,\infty)$, suppose for contradiction that the slow sampling instants accumulate at some finite $t_\infty$. Choosing $T=t_\infty+1$ leads to a contradiction because on $[0,T]$ the inter-event times are lower bounded by $\tau^*(T)>0$. 
Hence, the mechanism \eqref{ETmechanismslow} is Zeno-free on $[0,\infty)$, i.e., $\lim_{k\to\infty}t_k^y=\infty$. 
A similar argument applied to the fast sampling error $d_{v,\varepsilon}(t)$ ensures the existence of $\tau_*(T)>0$ such that $|s^y_{j+1}-s^y_j|\ge \tau_*(T)$ for all $j$ with $\varepsilon s_{j+1}^{y}\leq T$, and consequently the fast triggering mechanism \eqref{ETmechanismfast} is also Zeno-free on $[0,\infty)$, i.e., $\lim_{j\to\infty}s_j^y=\infty$.

\bibliographystyle{IEEEtranPF}
\bibliography{ref}

\end{document}